\documentclass[aps,superscriptaddress,showpacs,nofootinbib,eqsecnum]{revtex4}

\input{epsf}

\begin{document}

\title{Higher Order Evaluation of the Critical Temperature for 
Interacting Homogeneous Dilute Bose Gases}

\author{Frederico F. de Souza Cruz}
\email{fred@fsc.ufsc.br}
\affiliation{Departamento de F\'{\i}sica,
Universidade Federal de Santa Catarina,
88040-900 Florian\'{o}polis, SC, Brazil}

\author{Marcus B. Pinto}
\email{marcus@lpm.univ-montp2.fr}\thanks{$^1$ Permanent address}
\affiliation{Departamento de F\'{\i}sica,
Universidade Federal de Santa Catarina,
88040-900 Florian\'{o}polis, SC, Brazil}
\affiliation{Laboratoire de Physique Math\'{e}matique et Th\'{e}orique - CNRS - UMR
5825 Universit\'{e} Montpellier II, France}

\author{Rudnei O. Ramos}
\email{rudnei@dft.if.uerj.br}\thanks{$^3$ Permanent address}
\affiliation{Departamento de F\'{\i}sica Te\'orica,
Universidade do Estado do Rio de Janeiro,
20550-013 Rio de Janeiro, RJ, Brazil}
\affiliation{Department of Physics and Astronomy, Dartmouth College,
Hanover, New Hampshire 03755-3528}

\author{Paulo Sena}
\email{psena@bon.matrix.com.br}\thanks{$^5$ Permanent address}
\affiliation{Departamento de F\'{\i}sica,
Universidade Federal de Santa Catarina,
88040-900 Florian\'{o}polis, SC, Brazil}
\affiliation{Universidade do Sul de Santa Catarina, Av. Jos\'{e} A. Moreira, 787, 
88704-900 Tubar\~{a}o, SC, Brazil}

%last change by: Rudnei 27.02.02
%ultima modif marcus:25.02.02
%ultima modif fred  :
%ultima mod paulo:

\begin{abstract}

We use the nonperturbative linear $\delta$ expansion method to evaluate
analytically the coefficients $c_1$ and $c_2^{\prime \prime}$ which
appear in the expansion for the transition temperature for a dilute,
homogeneous, three dimensional Bose gas given by $T_c= T_0 \{ 1 + c_1 a
n^{1/3} + [ c_2^{\prime} \ln(a n^{1/3}) +c_2^{\prime \prime} ] a^2
n^{2/3} + {\cal O} (a^3 n)\}$, where $T_0$ is the result for an ideal
gas, $a$ is the s-wave scattering length and $n$ is the number density.
In a previous work the same method has been used to evaluate $c_1$ to
order-$\delta^2$ with the result $c_1= 3.06$. Here, we push the
calculation to the next two orders obtaining $c_1=2.45$ at
order-$\delta^3$ and $c_1=1.48$ at order-$\delta^4$. Analysing the
topology of the graphs involved we discuss how our results
relate to  other
nonperturbative analytical methods such as the self-consistent resummation and
the $1/N$ approximations. At the same orders we obtain $c_2^{\prime
\prime}=101.4$, $c_2^{\prime \prime}=98.2$ and $c_2^{\prime
\prime}=82.9$. Our analytical results seem to support the recent Monte
Carlo estimates $c_1=1.32 \pm 0.02 $ and $c_2^{\prime \prime}= 75.7 \pm
0.4$.

%PACS number(s):  03.75.Fi, 05.30.Jp, 11.10.Wx

\end{abstract}

\pacs{03.75.Fi, 05.30.Jp, 11.10.Wx}

\vspace{0.75cm}
\centerline{\it In press Physical Review A (2002)}

\maketitle

\section{Introduction}

Recently, the evaluation of the critical temperature for interacting
dilute homogeneous Bose gases has been the interest of many theoretical
works. {}For this purpose, the starting model is the
one used in the analysis of  a  gas of
interacting boson particles, described by a complex scalar field $\psi$,
with a local interaction characterized by the s-wave scattering length $a$
and Euclidean action which, in natural unities, can be written as

\begin{equation}
S_E = \int_0^\beta d\tau \int d^3 x \left\{
\psi^*({\bf x},\tau)\left(
\frac{d}{d\tau}-\frac{1}{2m}\nabla^2\right)
\psi ({\bf x},\tau)
-\mu \psi^* ({\bf x},\tau) \psi ({\bf x},\tau)
+  \frac{2 \pi a}{m}  \left[\psi({\bf x},\tau)\psi^*({\bf x},\tau)
\right]^2 \right\}\;.
\label{SE}
\end{equation}

The field
$\psi$ can be decomposed into imaginary-time frequency modes 
$\psi_j ({\bf x},\omega_j)$,
with discrete Matsubara frequencies $\omega_j = 2 \pi j/ \beta$ and
$j$ being an integer whereas $\beta$ is the inverse of the temperature.
At the early stages of solving this problem \cite{baymprl} the
non-zero Matsubara frequency modes have been integrated out generating
a reduced three dimensional $O(2)$ scalar theory. This procedure was
justified on the grounds that near the transition the non-zero Matsubara
modes decouple and one is left with an effective action given by

\begin{equation}
S_{3d} = \beta \int d^3 x \left\{
\psi_0^*\left(-\frac{1}{2m}\nabla^2- \mu \right )
\psi_0 + \frac{2 \pi a}{m}  \left[\psi_0 \psi_0^*
\right]^2 \right\}\;.
\label{3d}
\end{equation}
Despite this simplification the problem remains non-trivial since
ordinary perturbation theory cannot be used to treat the model at the
phase transition due to the severe infrared divergences for the zero
frequency modes $\psi_0$ at the critical point, originating the breakdown of
conventional perturbation theory. Different nonperturbative
methods, some of which are currently used in
quantum field theories, have then been used to compute the transition
temperature. The analytical methods include the self-consistent
resummation (SCR) used by the authors of Ref. \cite {baymprl},
the $1/N$ expansion used at leading order ($1/N$-LO)  by Baym, Blaizot and
Zinn-Justin \cite {baymN} and at next to leading order ($1/N$-NLO) by Arnold and
Tom\'{a}sik \cite {arnold} as well as the linear $\delta$-expansion
(LDE) employed by some of the present authors in Ref. \cite {prb}. The numerical
methods used mainly Monte Carlo lattice simulations (MCLS) like the ones employed
recently by
Arnold and Moore \cite {arnold1} and by Karshunikov, Prokof'ev and
Svistunov \cite {russos}. Most of those calculations predicted that, in
the dilute limit, the shift of the critical temperature of the interacting
gas, $T_c$, as compared to the critical temperature for an ideal gas,
$T_0$, $\Delta T_c=
T_c-T_0$, behaves as

\begin{equation}
\frac { \Delta T_c} {T_0} =  c_1 a n^{1/3} + {\cal O} \left(a^2 n^{2/3}
\right) \;,
\label{tc}
\end{equation}
where $n$ is the number density, $c_1$ is a numerical constant and
the critical temperature for an ideal gas is given as usual by

\begin{equation}
T_0 =  \frac{2 \pi}{ m} \left[\frac{n}{\zeta(3/2)}
\right]^\frac{2}{3}\;.
\label{T0}
\end{equation}

The constant $c_1$ in Eq. (\ref{tc}) is directly related to the
contributions from the zero mode Matsubara frequencies and therefore can
only be computed from nonperturbative methods. Some recent numerical
applications predicted values for $c_1$ which are close to $1.30$ (MCLS,
\cite{arnold1,russos}). On the other hand, the analytical applications
mentioned above predicted the values $2.90$ (SCR, \cite {baymprl}),
$2.33$ ($1/N$-LO, \cite {baymN}), $1.71$ ($1/N$-NLO, \cite {arnold}) and
$3.06$ (LDE, \cite {prb}). Additionally, the authors of Ref.
\cite{gordon} have also argued that a logarithmic term appears at
order-$a^2$ in Eq. (\ref {tc}). They have shown that this term is of the
form $ c_2^{\prime} a^2 n^{2/3} \ln(a n^{1/3})$ and also estimated,
using large-$N$ arguments, the value of the numerical coefficient
$c_2^{\prime}$. Recently, Arnold, Moore and Tom\'{a}sik \cite {second}
have argued that when naively going from the original action ($S_{E}$)
to the reduced action ($S_{3d}$) by ignoring the effects of non-zero
frequency modes one misses the effects that short-distances and/or
high-frequency modes have on long-distance physics. {}For $T_c(n)$ at
second order these effects can be absorbed into a modification of the
strengths of the relevant interactions which means that one should consider the more
general form for the reduced effective action Eq. (\ref{3d})

\begin{equation}
S_{\rm eff} [\psi_0,\psi^*_0] = \beta \int d^3 x \left\{\psi^*_0 \left(
-{\cal Z}_{\psi} \frac{1}{2m}\nabla^2 -\mu_{3} \right)
\psi_0
+{\cal Z}_a \frac{2 \pi a}{m}  \left[\psi^*_0\psi_0
\right]^2 + {\cal O}\left[ \psi^*_0 \psi_0 |\nabla\psi|^2, (\psi^*\psi)^3\right]
\right\} + \beta F_{\rm vacuum}\;,
\label{Seff}
\end{equation}

\noindent
where ${\cal Z}_\psi$ is the wave-function normalization
function, $\mu_3$ incorporates the mass renormalization function,
${\cal Z}_a$ incorporates the vertex renormalization function and
$F_{\rm vacuum}$ represents the vacuum energy contributions coming from
the integration over the nonstatic Matsubara modes. The ${\cal
O}\left[ \psi^*_0 \psi_0 |\nabla\psi_0|^2, (\psi_0^*\psi_0)^3\right]$
terms represent higher order interactions in the zero modes of the fields. As
emphasized in Ref. \cite{second}, these terms
will give contributions to the density of order $a^3$ and higher and
therefore do not enter in the order-$a^2$ calculations.
By matching perturbative order-$a^2$ results
obtained with the original action $S_{E}$ and the general effective
action $S_{\rm eff}$, the authors of  Ref. \cite{second} were able to 
show that the transition temperature for
a dilute, homogeneous, three dimensional Bose gas can be expressed at
next to leading order as

\begin{equation}
\frac { \Delta T_c} {T_0} =  c_1 a n^{1/3} + \left[ c_2^{\prime} \ln(a n^{1/3})
+c_2^{\prime \prime} \right] a^2 n^{2/3} + {\cal O} \left(a^3 n \right)\;.
\label {exptc}
\end{equation}
A similar structure is also discussed in Ref. \cite {markus}.
As far the numerical coefficients are concerned, the {\it exact} value for 
$c_2^{\prime}$, $c_2^{\prime}=-64 \pi \zeta(1/2)
\zeta(3/2)^{-5/3}/3 \simeq 19.7518$,
was obtained using perturbation theory \cite {second}. The other two coefficients
cannot be obtained perturbatively but they can, through the matching
calculation, be expressed in terms of the two nonperturbative
quantities $\kappa$ and ${\cal R}$ which are, respectively, related to
the number density $\langle \psi_0^* \psi_0 \rangle$ and to the critical
chemical potential $\mu_c$, as shown below. 
The actual relation in between the two nonperturbative 
coefficients and these physical quantities is given by
\cite {second}

\begin{equation}
c_1 = - 128 \pi^3  [\zeta(3/2)]^{-4/3}  \kappa   \;\;,
\label{c1}
\end{equation}
and

\begin{equation}
c_2^{\prime \prime} = - \frac{2}{3} [\zeta (3/2)]^{-5/3} b_2^{\prime \prime} + 
\frac {7}{9} [\zeta (3/2)]^{-8/3} (192 \pi^3 \kappa)^2 + \frac{64 \pi}{9} 
\zeta (1/2) [\zeta(3/2)]^{-5/3}
\ln \zeta (3/2)   \;,
\label{c2}
\end{equation}
where $b_2^{\prime \prime}$ in Eq. (\ref{c2}) is given by

\begin{equation}
b_2^{\prime \prime} = 32 \pi \left \{ \left [ \frac{1}{2} \ln (128 \pi^3) + 
\frac{1}{2} - 72 \pi^2 {\cal R} - 96 \pi^2 \kappa \right ]\zeta(1/2)
 + \frac {\sqrt {\pi}}{2} - K_2 - \frac {\ln 2}{2 \sqrt {\pi}}\left [ 
\zeta(1/2) \right ]^2 \right \}  \; ,
\label{b2primeprime}
\end{equation}
with $K_2= -0.13508335373$. The quantities $\kappa$ and ${\cal R}$ are related
to the zero Matsubara modes only. Therefore, they can be nonperturbatively 
computed directly from the reduced action
$S_{\rm eff}$ which, as discussed in the numerous previous
applications, can be written as

\begin{equation}
S_{\phi}=  \int d^3x \left [ \frac {1}{2} | \nabla \phi |^2 +
\frac {1}{2} r_{\rm bare}
\phi^2 + \frac {u}{4!} (\phi^2)^2
\right ] \;,
\label{action2}
\end{equation}

\noindent
where $\phi =  (\phi_1, \phi_2)$ is related to the original real
components of $\psi_0$ by $\psi_0({\bf x}) =\sqrt{mT/{\cal Z_\psi}}\;
[\phi_1({\bf x})+i\phi_2({\bf x})]$,  
$r_{\rm bare}=-2m \mu_{3}/{\cal Z_\psi}$ and
$u=48 \pi a mT ({\cal Z}_a/{\cal Z_\psi}^2$).
The vacuum contribution
appearing in Eq. (\ref{Seff}) will not enter in the specific
calculation we do here.

The three dimensional effective
theory described by Eq. (\ref {action2}) is super renormalizable
\footnote {Recall that the coupling constant, $u$, has dimensions of
mass in natural unities.} requiring only a mass counterterm  to eliminate any 
ultraviolet divergence.
In terms of Eq. (\ref{action2}), the
quantities $\kappa$ and ${\cal R}$ appearing
in Eqs. (\ref{c1}) - (\ref{b2primeprime}) are defined by \cite{second}

\begin{equation}
\kappa \equiv \frac {\Delta \langle \phi^2 \rangle_c}{u}  =
\frac { \langle \phi^2 \rangle_u - \langle \phi^2 \rangle_0}{u} \;,
\label{kappa}
\end{equation}
and 

\begin{equation}
{\cal R} \equiv \frac {r_c}{u^2} = -\frac {\Sigma(0)}{u^2}  \;,
\label{R}
\end{equation}
where the subscripts $u$ and $0$ in Eq. (\ref{kappa}) 
mean that the density is to be
evaluated in the presence of interactions and in the absence of
interactions, respectively, and 
$\Sigma(0)$ is the self-energy with zero external momentum.
Since they dependent on the zero modes
their evaluation is valid, at the critical point, only when done in a 
nonperturbative fashion.
As discussed in the next section the relation between $r_c$ and
$\Sigma(0)$ comes from the Hugenholtz-Pines theorem
at the critical point. 

Eq. (\ref{exptc}) is a general order-$a^2$ result with coefficients
that, therefore, depend on nonperturbative physics via $\kappa$ and
${\cal R}$. In principle, to evaluate these two quantities one may start
from the effective three-dimensional theory, given by Eq. (\ref
{action2}), and then employ any nonperturbative analytical or numerical
technique. In general, the analytical nonperturbative methods give a
prescription so as to select and sum an {\it infinite} number of
contributions belonging to a given class. {}For example, the infinite
subset that contains only direct (tadpole) contributions represents the
Hartree approximation, whereas exchange contributions are also taken
into account in the Hartree-Fock approximation. In practice the sum is
achieved by using a modified (``dressed") propagator to evaluate
physical quantities. The nonperturbative results are then generated by
solving self-consistent equations. However, in resumming calculations
the bookkeeping and renormalization may become a problem beyond leading
orders. 

Another popular analytical nonperturbative technique is the $1/N$
expansion \cite{zj,coleman} where one sums infinite subsets of
contributions whose order is labeled by ${\cal O}(1/N^n)$ where $N$ is
the number of field components. In general, the leading order
contribution is easily evaluated and may reveal interesting
nonperturbative physics, at least from a qualitative point of view,
apart for providing an ``exact" result within the large-$N$ limit. A
nice illustration is provided by its application, for example, to the
Gross-Neveu model at zero temperature, where the issue of chiral
symmetry breaking as well as asymptotic freedom were investigated
\cite{gn}. {}From a quantitative point of view the leading order may not
be sufficient and lead to errors since $N$ is finite and not too large
in most cases. An example of this case is illustrated by treating the
same Gross-Neveu model at finite temperature, where the leading order
large-$N$ calculation predicts a finite value for the critical
temperature at which chiral symmetry restoration takes place, in
contradiction to Landau's theorem for phase transitions in one space
dimension \cite{rose}. 

In practice, going to higher orders can be
a difficult task. Nevertheless,
the $1/N$ ranks as a good method to investigate nonperturbative physics
as shown in many applications. In particular, the results provided by
this approximation for the interacting Bose gas case, where $N=2$, are
surprisingly good already at leading order \cite{baymN}. Good numerical
results can also be obtained with self-consistent methods despite
some potential problems as discussed in Ref. \cite {markus}. The
numerical calculations use mainly Monte Carlo lattice techniques and
many different results, for the interacting Bose gas critical
temperature problem, were generated in this way. The differences arise
mainly from the way the theory is put on the lattice, the size of the
lattice, the way the continuum
limit is taken and other issues.
As already mentioned, two recent works seem to have settled this
question \cite {arnold1,russos}.

Here we shall present, and then apply, an alternative analytical
nonperturbative method known as the linear $\delta$ expansion (LDE)
\cite{linear,duncan}
(for earlier works, see for instance, Ref. \cite{early}), which 
is closely related to the variational perturbation
theory \cite{kleinert} and  the Gaussian effective potential \cite {pms}.
 This same method re-appeared under the name of optimized perturbation
theory \cite {hatsuda}. The main attractive feature of
this approximation is the fact that the actual evaluation of a physical
quantity, including the selection of the {\it finite} subset of relevant
contributions at each order, is done exactly as in perturbation theory.
It is then easy to control and explicitly evaluate one by one each of
the reduced number of contributions appearing at each order. The
implementation of the renormalization procedure follows the one
performed in most quantum field theory textbooks \cite {ramond}. After the
usual perturbative manipulation one generates nonperturbative results
through an optimization procedure, as we will discuss in the next
section.

This work is organized as follows. In Sec. II
we present the method and illustrate it with a simple application to
the pure anharmonic oscillator. In the same section we implement the
method in the effective three dimensional theory given by Eq. (\ref{action2})
in order to evaluate the constants $\kappa$ and ${\cal R}$,
Eqs. (\ref{kappa}) and (\ref{R}). The
quantity $r_c$ is then evaluated in Sec. III whereas $\langle \phi^2
\rangle$ is evaluated in Sec. IV. The optimization procedure is
carried out in Sec. V where the numerical results are presented and
compared with some of the recent results. We present our conclusions in
Sec. VI. All contributions, which include difficult five-loop {}Feynman
diagrams with arbitrary $N$, are explicitly evaluated by brute force 
without recurring to
any approximations. An appendix is included to show the details of the 
calculations of these higher order terms. To our knowledge, some of them 
have not been evaluated in this way before.

\section {The method and its application to the interacting Bose gas problem}

\subsection {The linear $\delta$ expansion}

The linear $\delta$ expansion (LDE) was conceived to treat nonperturbative 
physics while staying within the familiar calculational
framework provided by perturbation theory. In practice, this can be
achieved as follows. Starting from an action $S$ one performs the
following interpolation

\begin{equation}
S \rightarrow S_{\delta} =  \delta S + (1 - \delta) S_0(\eta) \;,
\label{s}
\end{equation}
which reminds the trick consisting of adding and subtracting a mass term
to the original action. One can readily see that at $\delta=1$ the
original theory is retrieved. This parameter is really just a bookkeeping
parameter and some authors do not even bother considering it
explicitly as we do \cite {phil}. The important modification is encoded
in the field dependent quadratic term $S_0(\eta)$ that, for dimensional
reasons, must include terms with mass dimensions ($\eta$). In principle,
one is free so as to choose these mass terms and within the Hartree
approximation they are replaced by a direct (or tadpole) type of 
self-energy before one performs any calculation. In the LDE they are taken as
being completely arbitrary mass parameters which will be fixed at the
very end of a particular evaluation. One then formally pretends that
$\delta$ labels interactions so that $S_0$ is absorbed in the propagator
whereas $\delta S_0$ is regarded as a quadratic interaction. So, one
sees that the physical essence of the method is the traditional dressing
of the propagator to be used in the evaluation of physical quantities
very much as in the Hartree case. What is different in between the two
methods is that with the LDE the propagator is completely arbitrary
while it is constrained to cope only with direct terms within the
Hartree approximation. So, within the latter approximation the relevant
contributions are selected according to their topology from the start.

Within the LDE one calculates in powers of $\delta$ as if it was small.
In this aspect the LDE resembles the large-$N$ calculation since both
methods use a bookkeeping parameter which is not a physical parameter
like the original coupling constants and within each method one performs the
calculations formally working as if $N \rightarrow \infty$ or $\delta
\rightarrow 0$, respectively. {}Finally, in both cases the bookkeeping
parameters are set to their original values at the end which, in our
case, means $\delta=1$. However, quantities evaluated with the LDE
dressed propagator will depend on $\eta$ unless one could perform a
calculation to all orders. Up to this stage the results remain strictly
perturbative and very similar to the ones which would be obtained via a
true perturbative calculation. It is now that the freedom in fixing
$\eta$ generates nonperturbative results. Since $\eta$ does not belong
to the original theory one requires that  a physical quantity $\Phi$
calculated with the LDE be evaluated at the point where it is less
sensitive to this parameter. This criterion, known as the Principle of
Minimal Sensitivity (PMS), translates into the variational relation \cite{pms}

\begin{equation}
\frac {d \Phi}{d \eta}\Big |_{\bar \eta} = 0 \;.
\label{pms}
\end{equation}
 
The optimum value $\bar \eta$ which satisfies Eq. (\ref{pms}) must be a
function of the original parameters including the couplings, which
generates the nonperturbative results. The convergence properties of
this method has been rigorously proved in the context of the anharmonic
oscillator (AO) \cite {phil,ian,kleinert2,guida}. Very recently, Kneur and Reynaud
\cite {damien} claimed to
have proved the convergence of this method in renormalizable quantum
field theories. These are very encouraging results for the present
application which uses a renormalizable effective model which shares
many similarities with the pure AO. Let us quickly illustrate how
this method works by considering the anharmonic oscillator described, in
Minkowski space, by

\begin{equation}
{\cal L} =\frac {1}{2} (\partial_0 \phi)^2 - \frac{1}{2} m^2 \phi^2 - 
\frac {\lambda}{4}\phi^4  \;.
\label {ao}
\end{equation}
If one sets $m=0$ in the relation above the model describes the pure
anharmonic oscillator which cannot be treated by usual perturbation
theory. Let us first consider the ground state energy density whose
exact result, $ {\cal E}^{\rm exact} = \lambda^{1/3} 0.420 804 974
478 \ldots$,  has been calculated by Bender, Olaussen and Wang \cite {carl}.
{}Following Eq. (\ref{s}) one may write the interpolated action as

\begin{equation}
{\cal L}_{\delta}= \frac {1}{2} (\partial_0 \phi)^2 - \frac{1}{2} 
\eta^2 \phi^2 - \delta \frac {\lambda}{4}\phi^4
+ \delta \frac{1}{2} \eta^2 \phi^2\;,
\label {aoint}
\end{equation}
from which one obtains the perturbative order-$\delta$ result  \cite {phil}

\begin{equation}
{\cal E}^{(1)}= -\frac{i}{2}\int_{-\infty}^{+\infty}
\frac {dp}{2\pi} \ln[p^2-\eta^2] -
\delta  \frac{i}{2}\int_{-\infty}^{+\infty} \frac {dp}{2\pi} 
\frac {\eta^2}{p^2-\eta^2} -
\delta \lambda \frac{3}{4}\left ( \int_{-\infty}^{+\infty} 
\frac {dp}{2\pi} \frac {1}{p^2-\eta^2} \right )^2 +{\cal O}(\delta^2) \;.
\end{equation}
Now, setting $\delta=1$ and applying the PMS optimization procedure one gets

\begin{equation}
{\bar \eta} = 3i \lambda  \int_{-\infty}^{+\infty} \frac {dp}{2\pi} 
\frac {1}{p^2-{\bar \eta}^2} \;,
\label{gap}
\end {equation}
which is a self-consistent mass gap equation. It can be easily checked that
with this solution one resumms exactly the same contributions that would
appear in the usual Hartree approximation. The same procedure will
capture the physics which arises from exchange terms at order-$\delta^2$
where the first contribution of this type appears together with
order-$\delta^2$ direct (Hartree) contributions. Moreover, as shown in
other applications \cite {gastao}, the result furnished by Eq.
(\ref {gap}) remains valid at second order if one considers only the
direct terms and this pattern is valid at any order in $\delta$. The
actual value predicted at this lowest order is $ {\cal E}^{(1)}={\cal
E}^{\rm Hartree} \sim \lambda^{1/3} 0.429$  which is
only about 2$\%$ greater than the exact result. As shown in Ref. \cite
{phil} this result can still be improved as one goes to higher orders. Here, we
shall be mainly concerned with the nonperturbative evaluation of the
vacuum expectation value $\langle \phi^2 \rangle$.
This quantity, whose exact result is $\langle
\phi^2 \rangle^{\rm exact} = \lambda^{-1/3} 0.456119955748 \ldots$ \cite
{baner}, was also evaluated  in Ref. \cite {phil}. The optimum values were
obtained with
$\bar \eta$ values coming from its direct optimization and also from the
optimization of ${\cal E}$. At order-$\delta$ the value $\langle \phi^2
\rangle^{(1)} = \lambda^{-1/3}0.446456$ was obtained from the direct
optimization ($\bar \eta = 1.259921$) and $\langle \phi^2 \rangle^{(1)}
= \lambda^{-1/3}0.436789$ was obtained from the injection of $\bar
\eta=1.14471$, which was generated by the optimization of ${\cal E}$.
One then sees that the optimum $\langle \phi^2 \rangle$ numerical values
generated by the two optimization procedures are very similar which could
be expected since, at each order, the diagrams which contribute to
$\langle \phi^2 \rangle$ and ${\cal E}$ have the same structure.

At this stage it should be clear how nonperturbative results may be
generated, through the variational PMS procedure, from the perturbative
evaluation of physical quantities. As already mentioned, the effective
model to be considered in the sequel for the description of the dilute
Bose gas temperature bears may similarities with the AO. The main
differences being the number of space-time dimensions concerning each
case (which means that one has to deal with ultraviolet divergences in
three dimensions) and the fact that the former is used to investigate a
phase transition. Technically, as we shall see, this translates into
extra difficulties due to the Hugenholtz-Pines theorem which washes out
direct (tadpole) contributions, meaning that the first non-trivial
contributions to $\langle \phi^2 \rangle$ start at the three-loop level
via two-loop self-energies. Apart from the quantum mechanical
applications \cite{phil,ian,kleinert2,guida}, the LDE was successfully applied to
the description of mesoscopic systems \cite{miles}, nuclear matter
properties \cite {gastao}, phase transitions in the scalar $\lambda \phi^4$
model \cite {prd1,hugh} as well as in the Gross-Neveu model \cite{su},
investigation of chiral symmetry phenomena in
QCD \cite{jloic} and in the determination of the
equation of state for the Ising model \cite {zjising}. It is worth
mentioning that the application of the LDE to the scalar $O(N) \times
O(N)$ model \cite {prd} has allowed to investigate the nonperturbative phenomenon
of symmetry nonrestoration at high temperatures further than it was
possible with other standard nonperturbative methods.

The first application of this method to
the present problem was performed in Ref. \cite {prb}, where only the
first non-trivial contribution, which appears at order-$\delta^2$, was
considered. A successful extension to the ultra-relativistic case was performed
by Bedingham and Evans in Ref. \cite{tim}.

\subsection {The interpolated theory for the zero frequency Matsubara modes}

One can now write the interpolated version of the effective model
described by Eq. (\ref {action2}). Before doing that let us rewrite
$r_{\rm bare} = r + A$ where $A$ is a mass counterterm coefficient.
This counterterm is  the
only one effectively needed within the modified minimal subtraction
({$\overline {\rm MS}$}) renormalization scheme which we will adopt
here. Then, one can choose

\begin{equation}
S_0 =  \frac{1}{2}  \left [ | \nabla \phi|^2 + \eta^2 \phi^2  \right ] \;,
\label{S0}
\end{equation}

obtaining

\begin{equation}
S_{\delta}=  \int d^3x \left [ \frac {1}{2} | \nabla \phi |^2 + 
\frac {1}{2} \eta^2 \phi^2  +\frac{\delta}{2}(r - \eta^2) \phi^2 +
\frac {\delta u}{4!} (\phi^2)^2   + \frac {\delta}{2} A_{\delta} \phi^2
\right ] \;\;.
\label{Sdelta}
\end{equation}

\noindent
Note that we have treated $r$ ($r_c$ at the critical point) as an interaction, 
since this
quantity has a critical value which is at least of order $\delta$. The
{}Feynman rules for this theory, in Euclidean space, are $-\delta r$,
$\delta \eta^2$ and $-\delta A_{\delta}$, for the quadratic vertices and 
$-\delta u$ for the quartic vertex. The propagator is given by

\begin{equation}
G^{(0)}(p)=[p^2+\eta^2]^{-1} \;.
\end{equation} 
The
corresponding diagrams for these rules are shown in {}Fig. 1. Note that
$\eta$ acts naturally as an infrared cutoff so we do not have to worry
about these type of divergences. By introducing only quadratic terms the
LDE interpolation does not alter the polynomial structure, and hence the
renormalizability, of the theory. 

In general, the counterterm
coefficients appearing in the interpolated theory have a trivial
dependence on the bookkeeping parameter and the
renormalization process can be consistently achieved with the
interpolated theory exactly as in ordinary perturbation theory. Once
inserted into a diagram, the extra quadratic vertex proportional to
$\delta \eta^2$ brings in more propagators decreasing the ultraviolet
degree of divergence. We point out that renormalization should be
carried out before the optimization process to ensure that the optimum value
$\bar \eta$ is a finite quantity. The interested reader is
referred to Refs. \cite {prd1,prd} for more details concerning renormalization
within the LDE.

Requiring that at the
critical temperature the original system must exhibit infinite
correlation length, means that, at $T_c$ and $\delta= 1$ (the
original theory), the full propagator $G^{(\delta)} (p)$, given by

\begin{equation}
G^{(\delta)}(p)=
\left[ p^2 + \eta^2 + \delta r -
\delta{\eta^2} + \Sigma^{(\delta)}_{\rm ren}(p)      \right  ]^{-1}\;,
\label {G}
\end{equation}
must satisfy $G^{(\delta)}(0)^{-1} =0$, which implies

\begin{equation}
\delta r^{(\delta)}_c = -\Sigma^{(\delta)}_{\rm ren}(0)\;.
\label{HP}
\end{equation}
The above equation is equivalent to the Hugenholtz-Pines theorem 
applied to the LDE.
The relation Eq. (\ref{HP}) shows that, to order-$\delta^{n}$,
the quantity $\delta r_c^{(n)}$ is directly obtained from the evaluation of
$\Sigma^{(n)}_{\rm ren}(0)$. As discussed in the introduction, we will
use the {}Feynman rules described above to evaluate perturbatively the
self-energy $ \Sigma^{(\delta)}_{\rm ren}(p)$ to order-$\delta^4$ from
which we will get the nonperturbative values for $\kappa$ and ${\cal R}$
by using the PMS optimization procedure. The subscript ``ren" in the
self-energy means that this quantity also contains all diagrams which
arise from the mass counterterm vertex proportional to $\delta A_{\delta}$.
{}For our purposes, the easiest way to obtain a perturbative expansion for
$\langle \phi^2 \rangle_u$ is to start from

\begin{equation}
\langle \phi^2 \rangle_u^{(\delta)}= \sum_{i= 1}^{N} 
\langle \phi^2_i \rangle_u^{(\delta)} =  N
\int \frac {d^3 p}{(2 \pi)^3} G^{(\delta)} (p) =
\int \frac {d^3 p}{(2 \pi)^3}
\frac{N}{p^2 +\eta^2}\left [ 1  + \frac {\delta(r^{(\delta)}_c - \eta^2) +
\Sigma^{(\delta)}_{\rm ren}(p) }{p^2 +\eta^2} \right ]^{-1} \;.
\label{p2}
\end{equation}

\noindent
Like $\delta r_c^{(n)}$, the order-$\delta^n$ quantity  
$\langle \phi^2 \rangle_u^{(\delta)}$ is obtained by evaluating the
self-energies to that order and subsequently  expanding the series on
the RHS of Eq. (\ref {p2}).
Therefore, to obtain
${\cal R}$ and $\kappa$ to order-$\delta^4$ we need to consider the fifty four 
self-energy contributions shown in {}Fig. 1.

\section{Evaluation of $r_c$ to ${\cal O}(\delta^4)$}

According to the Hugenholtz-Pines theorem, $\delta r_c^{(4)}$ is
obtained from the evaluation of all diagrams shown in {}Fig. 1 with zero
external momentum. To make this paper more pedagogical, let us do a step
by step evaluation of $r_c$ up to order-$\delta^2$. To order-$\delta$ one has
only the tadpole contribution, a direct application of the {}Feynman rules
for the interpolated theory and dimensional regularization (see appendix
for more details) gives the finite contribution

\begin{equation}
-\delta r_c^{(1)}= \Sigma^{(1)}_{\rm ren}(0) = - \delta 
u \frac{\eta}{8\pi} \left( \frac {N+2}{3} \right)  \;.
\end{equation}
Carrying on to order $\delta^2$ one considers the contributions depicted by
the first five diagrams of {}Fig. 1, which give

\begin{eqnarray}
-\delta r_c^{(2)}&=&  \Sigma^{(2)}_{\rm ren}(0) =  -\delta 
u \frac {\eta}{8\pi} \left ( \frac {N+2}{3} \right ) +
\delta^2 u \frac {\eta}{16\pi} \left ( \frac {N+2}{3} \right )  
- \delta^2 \frac {u }{16\pi} \frac{r_c}{\eta} \left(\frac {N+2}{3} \right)
\nonumber \\
&+& \delta^2 \frac {u^2}{128 \pi} \left ( \frac {N+2}{3} \right )^2   
- \delta^2 \frac{u^2}{(8\pi)^2} \frac{(N+2)}{18} \left[
\frac {1}{\epsilon} + 4 \ln \left( \frac{M}{\eta} \right ) - 2.394  \right] + 
\delta A_{\delta} + {\cal O}(\delta^3)  \;,
\label{rc2}
\end{eqnarray}
where $M$ is an arbitrary {$\overline {\rm MS}$} mass scale\footnote{
Note that our choice for
the integral measure, Eq. (\ref {rdim}), has already taken care of
constants like $4
\pi$ and $e^{\gamma_E}$ which, otherwise, would appear in the setting
sun logarithmic term.}. Now, one replaces $\delta r_c$
which appears at the right hand side with the value $\delta r_c^{(1)}$
obtained at the previous order so that the right hand side {\it remains}
of order $\delta^2$. Next, one sees that the setting sun, whose explicit
evaluation follows those performed in the appendix [see Eq.(\ref{ss0})],
displays an ultraviolet pole as $\epsilon \rightarrow \infty$. In fact,
within dimensional regularization, the only primitive ultraviolet
divergence associated with the effective super-renormalizable three
dimensional theory steams from the setting sun type of diagram with
three internal propagators. The pole associated with this divergence
fixes the mass counterterm coefficient in the modified minimal
subtraction renormalization scheme,
 
\begin{equation}
\delta A_{\delta} = \delta^2 \frac{u^2}{(8\pi)^2} \frac{(N+2)}{18} 
\frac {1}{\epsilon} \;.
\end{equation}
As usual, this ``vertex" must be considered also at higher orders (see
{}Fig. 1) so diagrams whose divergences arise from ``setting sun"
sub-diagrams may be rendered finite \cite {ramond}. Now, it is easy to
see how the ``double scoop" contribution (fourth term on the RHS of Eq.
(\ref {rc2})) is exactly canceled due to the HP condition applied to
$r_c$ at first order. One then gets the finite second order result

\begin{equation}
-\delta r_c^{(2)}=\Sigma^{(2)}_{\rm ren}(0)= -\delta u \frac{\eta}{8\pi} 
\left ( \frac {N+2}{3} \right )+
\delta^2 u \frac {\eta}{16\pi} 
\left ( \frac {N+2}{3} \right )
- \delta^2 \frac{u^2}{(8\pi)^2} \frac{(N+2)}{18} \left [
4 \ln \left ( \frac{M}{\eta} \right ) - 2.394  \right ]  + 
{\cal O}(\delta^3)
\end{equation}
Also at higher orders many contributions cancel. In special, any diagram
with one or more tadpole sub-diagram(s), like the ``double scoop"
discussed at order-$\delta^2$, disappear. Then, the diagrams which really
contribute to $\delta r_c^{(4)}$ are those shown in {}Fig. 2, where one
must consider the external lines as carrying zero momentum. At the same
time, counterterm diagrams associated with the zero external momentum
setting sun diagram (or sub-diagrams) could have been suppressed from that
figure. However, we prefer to write them explicitly so that the same
figure can be used again, facilitating the discussion in the next
section. Using the results obtained in the appendix one obtains the fourth order
result

\begin{eqnarray}
-\delta r_c^{(4)} &=& \Sigma^{(4)}_{\rm ren}(0)=   -
\delta u \frac{\eta}{8 \pi}\left ( \frac{N+2}{3} \right )
+\delta^2 u \frac{\eta}{16 \pi}\left ( \frac{N+2}{3} \right )
-\delta^2 \frac{u^2}{(4\pi)^2} \frac{(N+2)}{18} \left [
\ln \left ( \frac{M}{\eta} \right ) - 0.59775  \right ]  \nonumber \\
&+&  \delta^3 u \frac{\eta}{64 \pi}
\left ( \frac{N+2}{3} \right )  -
\delta^3 \frac {u^3}{\eta} \frac {(N+2)^2}{108( 4
\pi)^3}[0.143848]  +\delta^3 \frac {u^3}{\eta}
\frac{\left( 16 + 10 N +  N^2 \right)}{(4\pi)^5 108}  [81.076]\nonumber \\
&-&
\delta^3 u^2 \frac {(N+2)}{18( 4\pi)^2}[0.498]
 +\delta^4 u \frac{\eta}{128 \pi}\left ( \frac{N+2}{3} \right )
-\delta^4 \frac {u^3}{\eta} \frac {(N+2)^2}{108(4\pi)^3} [0.0610]
\nonumber \\
&+&\delta^4 \frac {u^4}{\eta^2} \frac {(N+2)}{6 (4\pi)^6}
\frac{\left( 16 + 10 N +  N^2 \right)}{108}    [8.09927]
-  \delta^4 \frac {u^3}{\eta} \frac {(N+2)^2}{108(4\pi)^3} [0.011788]
-  \delta^4 u^2 \frac {(N+2)}{18(4\pi)^2} [0.166492] \nonumber \\
&-&\delta^4 u^2 \frac {(N+2)}{18(4\pi)^2} [0.0834]
+\delta^4 \frac {u^3}{\eta}
\frac{\left( 16 + 10 N +  N^2 \right)}{(4\pi)^5 108}   [10.240] +\delta^4
\frac {u^3}{\eta}
\frac{\left( 16 + 10 N +  N^2 \right)}{(4\pi)^5 108}   [30.31096] \nonumber \\
&-&
\delta^4 \frac {u^4}{\eta^2}
\frac {\left( 40 + 32 N +8 N^2 + N^3 \right) }{(4\pi)^6 648}  [20.43048]
- \delta^4 \frac {u^4}{\eta^2}
\frac {\left( 44 + 32 N + 5 N^2 \right) }{(4\pi)^6 324}  [12.04114]
\nonumber \\
&-&
\delta^4 \frac {u^4}{\eta^2}
\frac { \left( 44 + 32 N + 5 N^2 \right) }{(4\pi)^6 324}
[17.00434]
+ \delta^4 \frac {u^4}{\eta^2} \frac {(N+2)^2}{(18)^2(4\pi)^6}
[2.8726]+  {\cal O}(\delta^5) \;.
\label {rc}
\end{eqnarray}
The scale dependence of this quantity will be discussed in Section V.

\section{Evaluation of $\langle \phi^2 \rangle_u$ to order $\delta^4$}

In principle, to obtain $\langle \phi^2 \rangle_u^{(4)}$ one should
consider all contributions to the self-energy $\Sigma^{(4)}_{\rm
ren}(p)$ given by the diagrams of {}Fig. 1 with external momentum $p$.
However, thanks to results of the previous section, one does not have to
do the evaluation of all those graphs explicitly at this stage. In
fact, one can immediately reduce the number of graphs to be considered
by substituting the vertex $\delta r$ with the the appropriate critical
value $\delta r_c$ obtained in the previous section. Then, as
for $\Sigma_{\rm ren }^{(4)}(0)$, the set of diagrams which effectively
contribute to $\Sigma^{(4)}_{\rm ren}(p)$ reduces to those shown in
{}Fig. 2, but now one must consider the external lines as carrying
momentum $p$. Substituting Eq. (\ref {rc}) into Eq. (\ref {p2})
one sees that the quantity which matters for the evaluation of $\langle
\phi^2 \rangle_u^{(4)}$ is $ \Sigma^{(4)}_{\rm ren}(p)-\Sigma^{(4)}_{\rm
ren}(0)$. Diagramatically, this quantity is given by taking the graphs
of {}Fig. 2 with zero external momentum and subtracting them from the
same diagrams with external momentum $p$. This means that all diagrams
which do not depend on the external momentum will not contribute
in the evaluation of $\langle \phi^2 \rangle$ at the critical point.
For example, all
the tadpole diagrams with any type of sub-diagrams will not contribute. 
As expected,
the mass counterterm is
a redundant quantity in the evaluation of $\langle \phi^2
\rangle_u^{(n)}$ because this quantity depends on the difference

\begin{equation}
\Sigma^{(n)}_{\rm ren}(p)-\Sigma^{(n)}_{\rm ren}(0)  = 
[ \Sigma^{(n)}_{\rm div}(p)+ \Sigma^{(n)}_{\rm ct}(p)]-
[\Sigma^{(n)}_{\rm div}(0)  + \Sigma^{(n)}_{\rm ct}(0) ] \;,
\end{equation}
where $\Sigma^{(n)}_{\rm div}(p)$ is the divergent self-energy. {}For a
general renormalizable theory, the quantity $\Sigma^{(n)}_{\rm ct}(p)$
represents all counterterms associated with the parameters of the theory
(such as masses and coupling constants) as well as the wave-function
counterterm associated with any eventual momentum dependent pole. At the
same time, $\Sigma^{(n)}_{\rm ct}(0)$ involves the same counterterms
except for the wave-function one. However, as we have already
emphasized, in the three-dimensional case the only type of primitive
divergence requires only a mass counterterm, which is the same for
$\Sigma^{(n)}_{\rm div}(p)$ and $\Sigma^{(n)}_{\rm div}(0)$. This means
that in our case, $\Sigma^{(n)}_{\rm div}(p)-\Sigma^{(n)}_{\rm div}(0)$
is always a finite quantity as shown explicitly in the appendix where
it is also shown that this quantity is scale independent, as opposed to
$r_c$. Therefore, the type of diagrams which really matter for the
evaluation of $\langle \phi^2 \rangle_u^{(4)}$ are those shown in {}Fig.
3, which can be obtained expanding Eq. (\ref{p2}) to ${\cal O}(\delta^4)$. 
{}Following the sequence of diagrams shown in {}Fig. 3 one can write

\begin{eqnarray}
\langle \phi^2 \rangle_u &=&  \int \frac {d^3 p}{(2 \pi)^3} 
\frac{N}{p^2 +\eta^2}\left \{1+ 
\frac{\delta\eta^2}{p^2 +\eta^2}+ \frac {\delta^2 \eta^4}{(p^2  
+\eta^2)^2}+ \frac {\delta^3 \eta^6}{(p^2  
+\eta^2)^3}+ \frac {\delta^4 \eta^8}{(p^2  
+\eta^2)^4}\right . \nonumber \\
&-& \delta^2\frac {[\Sigma_{1}(p) - \Sigma_{1}(0)]}{p^2 +\eta^2}
-\delta^3  \frac {2\eta^2[\Sigma_{1}(p) - \Sigma_{1}(0)]}{(p^2 +\eta^2)^2}
-\delta^3 \frac {[\Sigma_{2}(p) - \Sigma_{2}(0)]}{p^2 +\eta^2}\nonumber \\
&-& \delta^3\frac {[\Sigma_{3}(p) - 
\Sigma_{3}(0)]}{p^2 +\eta^2}-\delta^4 \frac {3\eta^4[\Sigma_{1}(p) - 
\Sigma_{1}(0)]}{(p^2 +\eta^2)^3} -
\delta^4  \frac {2\eta^2[\Sigma_{2}(p) - \Sigma_{2}(0)]}{(p^2 +\eta^2)^2} 
\nonumber \\
&-&\delta^4 \frac {[\Sigma_{4}(p) - \Sigma_{4}(0)]}{(p^2 +\eta^2)}
-\delta^4 \frac {[\Sigma_{7}(p) - \Sigma_{7}(0)]}{(p^2 +\eta^2)}
-\delta^4 \frac {2\eta^2[\Sigma_{3}(p) - \Sigma_{3}(0)]}{(p^2 +\eta^2)^2}
\nonumber \\
&-& \delta^4 \frac {[\Sigma_{10}(p) - \Sigma_{10}(0)]}{(p^2 +\eta^2)}
-\delta^4 \frac {[\Sigma_{5}(p) - \Sigma_{5}(0)]}{(p^2 +\eta^2)}
+\delta^4 \frac {[\Sigma_{1}(p) - \Sigma_{1}(0)]^2}{(p^2 +\eta^2)^2} 
\nonumber \\
&-&\delta^4 \frac {[\Sigma_{6}(p) - \Sigma_{6}(0)]}{(p^2 +\eta^2)}
-\delta^4 \frac {[\Sigma_{8}(p) - \Sigma_{8}(0)]}{(p^2 +\eta^2)}
-\delta^4 \frac {[\Sigma_{9}(p) - \Sigma_{9}(0)]}{(p^2 +\eta^2)} 
\nonumber \\
&-& \left . \delta^4 \frac {[\Sigma_{11}(p) - \Sigma_{11}(0)]}{(p^2 +\eta^2)}
+ {\cal O}(\delta^5) \right \}\;.
\label{expansion}
\end{eqnarray}

\noindent
The details of the explicit evaluation of the $\Sigma_i$ terms
are given in the appendix. The final result we obtain is

\begin{eqnarray}
\langle \phi^2 \rangle_u &=&
- \frac{N\eta}{4\pi}+\frac {\delta}{2} \frac{N\eta}{4\pi} + 
\frac{\delta^2}{8}\frac{N\eta}{4\pi} + \frac {\delta^3}{16} 
\frac{N\eta}{4\pi} +  \frac {\delta^4}{128} \frac{5 N\eta}{4\pi}
\nonumber \\
&-& {\delta^2} \frac {u^2}{\eta} \frac {N(N+2)}{18(4\pi)^3} [0.143848]- 
\delta^3 \frac {u^2}{\eta} \frac {N(N+2)}{18(4\pi)^3}[0.01168]    -
\delta^3 \frac {u^2}{\eta} \frac {N(N+2)}{18(4\pi)^3}
[0.0610] \nonumber \\
&+& \delta^3 \frac {u^3}{\eta^2} \frac{N}{(4\pi)^6}
\frac{\left(16+10 N + N^2 \right)}{108} [8.09927] -
\delta^4 \frac{u^2}{\eta}\frac {N(N+2)}{18(4 \pi)^3}  [2.8270 \times 10^{-3}]  
\nonumber \\
&-& \delta^4  \frac{u^2}{\eta}\frac {N(N+2)}{18(4\pi)^3}
[7.7318\times 10^{-3}] -   \delta^4 \frac{u^2}{\eta} \frac {N(N+2)}{18(4\pi)^3} 
 [0.02461]   - \delta^4 \frac{u^2}{\eta} 
\frac {N(N+2)}{18(4\pi)^3}   [0.01825]
\nonumber \\
&+&\delta^4 \frac {u^3}{\eta^2} \frac{N}{(4\pi)^6}
\frac{\left(16+10 N + N^2 \right)}{108} [0.85984]
+ \delta^4 \frac {u^3}{\eta^2} \frac{N}{(4\pi)^6}
\frac{\left(16+10 N + N^2 \right)}{108} [1.937786]
\nonumber \\
&+& \delta^4  \frac {u^3}{\eta^2} \frac{N}{(4\pi)^6}
\frac{\left(16+10 N + N^2 \right)}{108} [5.30476] - \delta^4
\frac {u^4}{\eta^3} \frac {N(N+2)^2}{ (18)^2 (4\pi)^7}[0.87339]
\nonumber \\
&-&\delta^4 \frac {u^4}{\eta^3} \frac{N}{(4 \pi)^7}
\frac{\left(40 + 32 N + 8 N^2 + N^3 \right)}{648} [3.15904767]   - \delta^4
\frac {u^4}{\eta^3} \frac{N}{(4 \pi)^7}
\frac {\left( 44 + 32 N + 5 N^2 \right)}{324} [1.70959]
\nonumber \\
&-& \delta^4
\frac {u^4}{\eta^3} \frac {N(N+2)^2}{(18)^2(4\pi)^7}[4.4411]
-\delta^4 \frac {u^4}{\eta^3} \frac{N}{(4 \pi)^7}
\frac{\left( 44 + 32 N + 5 N^2 \right)}{324}
[2.37741] +  {\cal O}(\delta^5) \;.
\label{first3}
\end{eqnarray}

\section{Numerical Results for the Temperature Shift}

In this section we will turn our, so far, perturbative evaluation into
nonperturbative results using the PMS optimization prescription. Our
analysis of results, including the selection of the relevant optima,
will follow closely those adopted in the applications which proved the
convergence of this method for the anharmonic oscillator
\cite{phil,ian,kleinert2,guida}. Some of the guidelines developed on those studies
are essential for our present application. Let us start the optimization
process with the scale independent quantity $\langle
\phi^2\rangle_u^{(\delta)}$ whose recent Monte Carlo estimate is
$\langle \phi^2\rangle_u= -0.001198(17)\,u$ \cite {second}.
Before optimizing let us remark that all contributions to
$\langle \phi^2\rangle_u^{(n)}$ are proportional to $\delta^n u^n
\eta^{1-n}$ and therefore, the PMS condition will imply solving a
polynomial equation of degree $n$. As one may expect, many of those $n$
roots which determine the optimum $\bar \eta$ will be complex. Also, as
observed in the anharmonic oscillator studies, most of the time the best
results are in fact generated by the complex solutions \cite {phil}. Since $\eta$ is
arbitrary we have no justification, a priori, to throw away its complex
part. This means that our optimized physical quantities $\langle
\phi^2\rangle_u$ and $r_c$ will have, eventually, complex parts whose
meaning is to be interpreted according to the physics. Here, these two
quantities are ultimately used to determine a strictly real physical
quantity defined by the critical temperature. Therefore, for our
purposes the complex parts of those two physical quantities are not
relevant and will not be considered. Note that the imaginary parts of
optimized physical observables have also been dropped in Ref. \cite {phil}
where a different, but still valid,
 physical argument has been used.
{}Finally, we shall follow the
original PMS prescription \cite {pms} and optimize $\langle
\phi^2\rangle_u^{(\delta)}$ and $r_c^{(\delta)}$ separately. This
procedure was also adopted in the ultra relativistic case where it has
produced good results \cite {tim}.

By truncating Eq. (\ref{first3}) to the
first nontrivial order, order-$\delta^2$, setting
$\delta= 1$ and by applying the PMS, one gets the two real roots

\begin{equation}
{\bar \eta} =\pm 0.0232332 \; u\;  \;,
\label{etabar}
\end{equation}
which give

\begin{equation}
\langle \phi^2\rangle_u^{(2)} =  \mp 0.002777326 \; u\;.
\label{result}
\end{equation}
Applying the PMS to $\langle \phi^2\rangle_u^{(\delta)}$ at
order-$\delta^3$ one obtains the following three solutions. The first, $\bar \eta =
-0.0475422 \,u$  gives $\langle \phi^2\rangle_u^{(3)} = 0.0045505
\, u$ while the other two

\begin{equation}
{\bar \eta} =  (0.0237711 \pm 0.0268995 i)\; u \;,
\label{etabar1}
\end{equation}
yield

\begin{equation}
\langle \phi^2\rangle_u^{(3)} = -(0.00221912 \pm 0.00150245 i)\, u \;.
\label{order3}
\end{equation}

At order-$\delta^4$ one obtains the real solutions ${\bar \eta} =
0.0439352 \, u$ which gives $ \langle \phi^2\rangle_u^{(4)}=
-0.00293974 \; u$ and ${\bar \eta} = -0.0697993 \, u$ which gives $
\langle \phi^2\rangle_u^{(4)}=0.00483554 \, u$. The complex solutions are

\begin{equation}
{\bar \eta} =  (0.0129321 \pm 0.04676942 i)\; u \;,
\label{etabar2}
\end{equation}
from which one gets

\begin{equation}
\langle \phi^2\rangle_u^{(4)} = -(0.00134323 \pm 0.00213104 i) \, u \;.
\label{order4}
\end{equation}

In order to select the appropriate roots we recur again to the AO
convergence studies where the existence and behavior of optima families
was fully investigated to order-$\delta^{47}$ \cite {phil}. There, it
was observed that at a given order $n$ each PMS solution belongs to a
different family, the exception being complex conjugate solutions which
belong to the same family. It was  observed that, in the complex
plane, the first member of a new family always lies on the real axis and
also that a new family arises as $n$ is increased by 2. Supposing that
these findings may also be used in our three-dimensional problem, we may
identify two families whose first members lie on the real axis at
order-$\delta^2$. {}Family 1 starts with the positive real solution ${\bar
\eta} = 0.0232332 \; u$ and family 2 with the negative real solution
${\bar \eta} =- 0.0232332 \; u$. No new families arise when one goes to
the next order and the real negative solution $\bar \eta = -0.0475422
\,u$ is just another member of the family of negative real solutions (2)
while the complex conjugate optima with positive real parts ${\bar \eta}
= (0.0237711 \pm 0.0268995 i)\; u$ are taken as belonging to family 1.
At order-$\delta^4$, family 2 gets another member given by ${\bar \eta}
= -0.06983 \, u$, whereas family 1 gets ${\bar \eta} = (0.0129321 \pm
0.04676942 i)\; u$.

As we have increased the order by 2 one effectively
sees the appearance of a new family whose first member lies on the real
axis and is given by ${\bar \eta} = 0.0439352 \, u$. We can now roughly
examine the convergence of our results. The values obtained with the
optima belonging to family 1 are $\langle \phi^2\rangle_u^{(2)}= -
0.002777326 \; u $, ${\rm Re}[ \langle \phi^2\rangle_u^{(3)}] = -
0.00221912 u$ and ${\rm Re}[\langle \phi^2\rangle_u^{(4)}] = -0.00134323
\;u$. {}Family 2 gives $\langle \phi^2\rangle_u^{(2)} = 0.002777326 \; u$, 
$\langle \phi^2\rangle_u^{(3)} = 0.00405505 \; u $ and
$\langle\phi^2\rangle_u^{(4)} = 0.00483554 \; u$, whereas family 3 gives
$ \langle \phi^2\rangle_u^{(4)}= -0.00293974 \; u$. Note that the first
$\langle\phi^2\rangle_u$ value predicted by family 3 is only about $5 \%$
greater than the first value predicted by family 1. It is very likely
that family 3 will become complex and, as for the AO, as we go to higher
orders families 1 and 3 will predict very similar values converging to
the exact value. {}Family 2, on the other hand, seems to have only real
components. It predicts values of $\langle\phi^2\rangle_u$ which
increase order by order with a sign which is opposite to the one
predicted by families 1 and 3. Moreover, in the AO, it was observed that
the complex families have better convergence behavior than the purely
real families. This analogy indicates that family 1 should produce
converging results.

We can justify pushing the analogy in between our effective three
dimensional model and its one dimensional version that far by remarking
that, at least to the order we consider here,
$\langle\phi^2\rangle_u^{(\delta)}$ can be expressed as a power expansion
of the form

\begin{equation}
\langle \phi^2 \rangle_u^{(4)} = N \sum_{i=0}^4 (-1)^{i+1}(u\delta)^i
[\eta (1-\delta)^{1/2}]^{1-i} B_i \, ,
\end{equation}
where $B_0 \sim 10^{-1}$, $B_1 = 0$, $B_2 \sim 10^{-5}$, $B_3 \sim
10^{-6}$ and $B_4 \sim 10^{-7}$.  This structure is similar to the one found in the one
dimensional case. This hints that both models may have similar convergence
properties making our procedure more legitimate.
It is also worth
pointing out that in our previous work, Ref. \cite{prb}, we had only the
order-$\delta^2$ result and it was not possible to do the same type of
comparison among the solutions to find an acceptable pattern of order by
order corrections. There, to choose among the two possible solutions,
$\langle \phi^2\rangle_u^{(2)} = \mp 0.002777326 \; u\;$, we had to use
different arguments and were also guided by results found with
other methods. By considering higher orders, as we have done here, we can
overcome this problem and the negative result,
$\langle \phi^2\rangle_u^{(2)} = - 0.002777326 \; u\;$, naturally appears as the
one which belongs to the most well behaved sequence of order by order
corrections.

We are now in position to evaluate $u\kappa =\Delta \langle \phi^2
\rangle_c^{(\delta)}$, so that $c_1$ can be determined via Eq. (\ref{c1}) with the
optima contained in family 1. As one could expect, $\bar \eta$ is always
proportional to $u$ since the latter quantity is the only quantity with
mass dimensions appearing in $\langle \phi^2\rangle_u^{(\delta)}$. This means that
the optimum value for the non-interacting vacuum expectation value
$\langle \phi^2\rangle_0^{(\delta)}$ will be zero at any order. This agrees with
the results of Ref. \cite {second}, where it was shown that this is
indeed the value obtained when the theory is regularized with
dimensional regularization. Then, $u \kappa =\Delta \langle \phi^2
\rangle_c^{(n)}= \langle \phi^2\rangle_u^{(n)}$ from which one finally obtains $c_1
= 3.06$, $c_1 = 2.45$ and $c_1 = 1.48$ at order $\delta^2$, $\delta^3$
and $\delta^4$, respectively\footnote{At this stage it should be clear
that it is preferable to optimize $\langle \phi^2 \rangle_u^{(\delta)}$
rather than $\Delta \langle \phi^2 \rangle^{(\delta)}$ because the
latter quantity is less $\eta$ dependent.}. These results are compared
with other analytical and numerical results in Table 1.

It is instructive to examine the topology of the diagrams
contributing at each order so that we can establish the links with other
nonperturbative methods. At second order the non-trivial contribution
arises from the setting sun (one plain bubble) type of diagram. At third
order one has, besides the setting suns with insertions, a new
contribution which arises from the two plain bubble type of diagram
(ninth graph shown in {}Fig. 3). However, this contribution, belongs
with the setting sun to a class of diagrams that would appear in a plain
bubble sum or in the leading order of a $1/N$ type of calculation. At
fourth order one considers again a three plain bubble contribution
(eighteenth diagram of {}Fig. 3) but more radical changes arise via
other type of vertex corrections like the correction to the plain bubble
that comes from the nineteenth and twentieth diagrams of {}Fig. 3.
{}Finally, the last diagram contains a different type of vertex
correction that would appear on a ladder type of summation. In fact, one
can easily evaluate which are the individual contributions of the
 five-loop diagrams shown in {}Fig. 3. The first of them gives a
contribution (in unities of $u^4/\eta^3$) of approximately $1.9 \times
10^{-9}$, the second gives $2.8 \times 10^{-8}$, the third $2.7 \times
10^{-8}$, the fourth $8.86 \times 10^{-9}$ and the fifth gives $2.6
\times 10^{-8}$. These numbers show that, at this order, the total
contribution from the ladder and bubble correction type of contributions
(third and fifth) are effectively twice that of the plain three bubble
one. 

It is also easy to see by drawing that the only corrections which
may appear at odd orders are those due to the {\it doubling} of a
bubble that already appeared at the previous order (increasing the 
``bubble chain").
At the same time, at
even orders, one is allowed to {\it insert} a new bubble anywhere
creating diagrams with completely different topologies. In other words,
in a perturbative expansion of $\langle \phi^2 \rangle_u^{(\delta)}$,
new topological classes of  graphs can arise only at even orders.

One can now
appreciate that the reason our order-$\delta^2$ result $c_1=3.06$ \cite {prb},
obtained by optimizing only one setting sun contribution, compares so
well with the value $c_1=2.90$, found by resumming setting sun
contributions in a self-consistent way \cite{baymprl}, is a 
consequence that 
both approximations consider the same type of diagrams.
On the other hand, when going
to order-$\delta^3$ one considers a new diagram but which, 
together with the setting sun, would also be
considered in a large-$N$ calculation. In the LED, its effect is to 
reduce the second order result to $c_1=2.45$.
Let us consider, for the moment, the only order-$\delta^4$ contribution 
that would also be
considered in a large-$N$ calculation. Graphically this contribution is
displayed by the second of the five-loop terms in {}Fig. 3. Not
surprisingly, we obtain the value $c_1=2.32$ which is very close to the
$c_1=2.33$ value obtained with the $1/N$ method at leading order \cite
{baymN} and the numerical differences may be due to the fact that
we have considered our symmetry factors in full, not only the highest
power of $N$. The four remaining five-loop contributions would be
considered in an $1/N$ type of calculation to the next order. Such a
calculation has been performed by Arnold and Tom\'{a}sik \cite {arnold} who
found $c_1=1.71$, which is approximately $27 \%$ smaller than the leading
order result. In our case this fact is confirmed  at
order-$\delta^4$, where the net effect of considering diagrams which
would belong to a next to leading order $1/N$ evaluation is to decrease
the value $c_1=2.32$ obtained with the graph that would appear at
leading order in the same approximation by roughly $35 \%$. As before
the numerical differences must be due to the full consideration of
powers of $N$ in each symmetry factor. It is not our aim to establish here a 
formal
relationship among the different approximations. Nevertheless, the discussion 
above can serve as a guide to understand how the LDE captures part of the 
nonperturbative
physics contained within the SCR and $1/N$ approximations.

In order to evaluate the coefficient $c_2^{\prime \prime}$ we now turn
to the optimization of the scale dependent $r_c$. Setting $\delta=1$ and
applying the PMS to $r_c^{(2)}$ generates one positive, real optimum
given by $\bar \eta = u/6 \pi$. It is important to note that this PMS
solution is a scale independent quantity. In fact, $r_c$ depends on the
($\overline{\rm MS}$) mass scale through the term proportional to $u^2
\ln (M/\eta)$ which appears in the order-$\delta^2$ setting sun term. It
is then easy to see that when this term is derived with respect to
$\eta$ the scale dependence automatically disappears turning our 
optimization procedure into a scale
independent process. As discussed below this situation will be verified at 
any order in $\delta$.

Next, in order
to get a numerical result for the optimized $r_c$ one must fix a scale and 
here we
choose $M=u/3$ which is the same scale \footnote {Our notation for the
mass scale ($M$) is different from the one used by the authors in Ref.
\cite{second} ($
\overline M$).} used by Arnold, Moore and Tom\'{a}sik in Ref. \cite {second}
where the
result found for this quantity is  $r_c (M=u/3) = 0.001920(2) \, u^2$.
The relation in between the values of $r_c^{(\delta)}$, evaluated at two different
($\overline{\rm MS}$) mass scales $M_1$ and
$M_2$, can be obtained from
Eq. (\ref {rc}) and reads

\begin{equation}
\frac {r_c^{(4)}(M_1)}{u^2} = \frac {r_c^{(4)}(M_2)}{u^2}  + 
\frac {(N+2)}{18(4\pi)^2} \ln \left (\frac {M_1}{M_2} \right )
\;.
\end{equation}
It is not too difficult to see that this relation will be verified at 
any order in $\delta$.
At order-$\delta^2$ the only diagram which is  scale dependent is the setting sun.
At a higher order ($n \ge 3$) this order-$\delta^2$ contribution can 
only appear as a subdiagram.
At the same order  a similar graph appears, but this time $\delta r$
replaces the setting sun insertion. However, the ``vertex"
$\delta r_c$ is always replaced (see Sec. III)  by its expansion in 
$\delta$ which contains, at order-$\delta^2$,
exactly the same scale dependent term as given by the setting sun, 
with a reversed sign. This means that,
apart from the order-$\delta^2$ setting sun, all contributions to
$\delta r_c^{(n)}$ are automatically scale independent.
Optimizing our order-$\delta^2$  result one gets

\begin{equation}
r_c^{(2)}= 0.00315236 \, u^2 \;\;.
\end{equation}
Proceeding to next order the PMS gives two complex solutions,
$\bar \eta = (0.0353678 \pm 0.0550091 i) u^2$  which yield

\begin{equation}
r_c^{(3)}= (0.00221321 \pm 0.00009661 i) \, u^2 \;\;.
\end{equation}
{}Finally, the order-$\delta^4$ optimization results are the real 
solution $\bar \eta = 0.0659334 \, u$, which yields
$ r_c^{(4)}= 0.00246153 \, u^2$ and the complex solutions
$\bar \eta = (0.00947463 \pm 0.0797262 i) u^2$, which generates

\begin{equation}
r_c^{(4)}= (0.00165411 \pm 0.000772567 i) \, u^2 \;\;.
\end{equation}
As in the previous case one sees that the first optima family starts with
a real value at order-$\delta^2$ and turns into a complex family at
order-$\delta^3$. At order-$\delta^4$ it receives a new complex member.
The first family generates the
real  values   $r_c^{(2)}= 0.00315236 \, u^2$, ${\rm
Re} [r_c^{(3)}]= 0.00221321 \, u^2$ and ${\rm Re}[r_c^{(4)}]= 0.00165411
\, u^2$ which are our selected values. Then, using Eq (\ref{c2})
together with the optima values obtained for $\kappa$ and ${\cal R}$ we
obtain, order after order, the results $c_2^{\prime \prime} =101.4$,
$c_2^{\prime \prime} =98.2$ and $c_2^{\prime \prime} =82.9$ for the
order-$a^2$ nonperturbative coefficient. As for $c_1$, these results
 compare well with the Monte Carlo estimate,
$c_2^{\prime \prime} =75.7 \pm 0.4$.

\section{Conclusions}

We have used the linear $\delta$ expansion to evaluate nonperturbatively 
the numerical coefficients appearing in the expansion for
the transition temperature for a dilute, homogeneous, three-dimensional
Bose gas given by $T_c= T_0 \{ 1 + c_1 a n^{1/3} + [ c_2^{\prime} \ln(a
n^{1/3}) +c_2^{\prime \prime} ] a^2 n^{2/3} + {\cal O} (a^3 n)\}$, where
$T_0$ is the result for an ideal gas, $a$ is the s-wave scattering
length and $n$ is the number density. This expansion for $T_c$
incorporates the effects of non-zero Matsubara modes \cite{second,markus}.
While the coefficient $ c_2^{\prime}$ has been exactly evaluated
using perturbation theory the question about the numerical values of the
other two coefficients, $c_1$ and $c_2^{\prime \prime}$ remains open and
has been the object of recent investigations. The reason behind
this difficulty is the fact that these coefficients can only be obtained
in a nonperturbative way. 

Due to the Hugenholtz-Pines theorem the first non-trivial contribution
appears at an order where one has to consider, at least, momentum
dependent two-loop self-energy diagrams. Considering higher order terms,
so as to get more accurate results, becomes rapidly difficult within the
existing nonperturbative methods as discussed in Ref. \cite {markus},
where the authors state that the complexity of the mathematical problem
does not allow a definitive prediction of the prefactor $c_1$, of the
term linear in $a$, from an analytic analysis. On the other hand, two
recent numerical results obtained with lattice simulations, which
predict $c_1 \sim 1.30$ \cite {arnold1,russos,second}, are being taken
very seriously. In a previous work, Ref. \cite{prb}, we have applied the
LDE to this problem obtaining the value $c_1 \sim 3.06$ at the first
non-trivial order ($\delta^2$). However, the quality of that
application was difficult to infer, from a quantitative point of view,
since only one approximant had been used. On the hand, the fact that at
order-$\delta^2$ with only one graph the optimization procedure was able
to generate a result numerically similar to the one obtained with a
self-consistent resummation (SCR) of two-loop momentum dependent
contributions \cite {baymprl} was encouraging. At that time, we were not
in position to elaborate any further about the convergence behavior of
that result. 

In the present work, we have again explicitly shown that the LDE method
offers, as its major advantage, the possibility to select, evaluate and
renormalize a physical quantity exactly as in the familiar perturbative
framework. Here, the contributions appearing at each order are not
selected according to their topology as within most nonperturbative
analytical cases. Contrary to some previous unfounded criticisms, no
uncontrolled errors arise in this type of perturbative calculation, most
notably in this application, where even the most cumbersome five-loop
contributions have been fully considered and evaluated without recurring
to any approximations as shown in the appendix. Another advantage is
that one does not have to worry about infrared divergences, since,
during the formal evaluation of graphs, the LDE arbitrary parameter
naturally acts as such before disappearing during the optimization
process. 
Also, the fact that a convergence proof for the quantum mechanical analogue of
the model considered here does exist \cite {phil,ian,kleinert2,guida} is an extra bonus.

At first one
could think that the multiplicity of possible real and complex
results generated by the
PMS constitutes the most serious disadvantage of the LDE. Nevertheless,
the quantum mechanical convergence studies of Ref. \cite {phil},
have shown how
meaningful nonperturbative physical results can still be obtained. As discussed in the text, those
studies have been crucial to our application for some important reasons like
showing how the possible solutions gather into real and complex families and emphasizing that better
results are generated by the complex ones.
We recall that, although different physical arguments have been used in each case, the imaginary parts
of the optimized physical observables generated by the complex families have also been dropped out in
Ref. \cite {phil}.
 As already mentioned,
our effective model displays the same
series structure for the physical observable
$\langle\phi^2\rangle_u^{(4)}$ as its quantum mechanical counterpart.
Taking all these facts into account we were able to obtain the results $c_1=3.06$,
$c_1=2.47$ and $c_1=1.48$ at second, third and fourth orders,
respectively. Our results approach, order after order, the   recent
Monte Carlo estimate, $c_1 \sim 1.3$.

Comparing our results and the
topology of the diagrams considered here with those belonging to the self-consistent
resummation of setting suns (SCR) and the $1/N$ approximation at leading
($1/N$-LO) and next to leading ($1/N$-NLO) orders we made clear that our
results are not a mere coincidence. In fact, the PMS is successively
chopping, order after order, nonperturbative information contained in
those approximations. Our results confirm the decrease in the value of
$c_1$ observed successively with the SCR, $1/N$-LO and $1/N$-NLO. The
numerical differences may be due to the fact that we do not make any
distinction among the different powers of $N$ which appear on the
symmetry factors since the LDE was envisaged to cope with arbitrary $N$.

We remark that a problem
regarding the sign of the coefficient $c_1$, which appeared in our previous
application, has disappeared at this higher order evaluation. We have also
investigated the quantity $r_c$ by evaluating all self-energy
contributions, with zero external momentum, up to order-$\delta^4$. Once
this quantity was optimized we have obtained the values $c_2^{\prime
\prime}=101.4$, $c_2^{\prime \prime}=98.2$ and $c_2^{\prime \prime}=82.9$ for
the next nonperturbative coefficient at second, third and fourth orders,
respectively. These results are in good numerical agreement with
the Monte Carlo result,  $c_2^{\prime \prime}=75.7$ \cite {second}.

In summary our analytical investigation seems to support, order by
order, the results obtained with other three analytical nonperturbative
methods. Our fourth-order numerical results compare well
with the recent results found in Refs.
\cite{arnold1,russos,second}. Additionally, there is an exciting possibility
that the method may offer a way of making a definitive analytical
prediction for the nonperturbative coefficients $c_1$ and $c_2^{\prime
\prime}$, which we are currently investigating.

%\acknowledgements
\begin{acknowledgments}

The
authors would like to thank Hagen Kleinert and Axel Pelster for their
help concerning the general $N$ symmetry factors and Philippe Garcia for
discussing the selection of the optima.
F.F.S.C., M.B.P. and R.O.R. were partially supported by Conselho
Nacional de Desenvolvimento Cient\'{\i}fico e Tecnol\'ogico
(CNPq-Brazil). R.O.R. also thanks partial support from the
``Mr. Tompkins Fund for Cosmology and Field Theory'' at
Dartmouth.
P.S. was partially supported by Associa\c{c}\~{a}o
Catarinense das Funda\c{c}\~{o}es Educacionais (ACAFE-Brazil).

\end{acknowledgments}

\appendix

\section{Evaluating the higher loop terms}

To make this work self-contained we shall outline, in this appendix, the
details of the explicit evaluation of all {}Feynman diagrams considered in
the evaluation of $\langle \phi^2 \rangle_u^{(4)}$ for arbitrary $N$.
We also remark 
that working out symmetry factors for many loop contributions with
generic $N$ is a problem on its own. Here, we have used the methods
developed by Kleinert's group in Berlin \cite {berlin}.

We regularize all diagrams with dimensional regularization in arbitrary
dimensions $d= 3-2\epsilon$ and carry the renormalization with the
modified minimal subtraction scheme ($\overline{\rm MS}$). So the momentum
integrals are replaced by

\begin{equation}
\int \frac {d^3 p}{(2 \pi)^3} \to \int_p   
\equiv \left(\frac{e^{\gamma_E} M^2}{4 \pi} \right)^\epsilon
\int \frac {d^d p}{(2 \pi)^d} \;,
\label{rdim}
\end{equation}

\noindent
where $M$ is an arbitrary mass scale and $\gamma_E \simeq 0.5772$ is the
Euler-Mascheroni constant. Very often, in evaluating the contributions
to $\langle \phi^2 \rangle_u$ one considers the integral

\begin{equation}
\int_p \frac{1}{(p^2 + \eta^2)^n} = \frac{\eta^{3 -2n}}{(4 \pi)^{3/2}} 
\frac {\Gamma[n+\epsilon -3/2]}{\Gamma(n)}
\left ( \frac {M^2 e^{ \gamma_E}}{\eta^2} \right )^{\epsilon} \;\;.
\label{intsimples}
\end{equation}
This integral can be explicitly evaluated as above or by considering
the case $n=1$

\begin{eqnarray}
\int_p \frac{1}{p^2 + \eta^2}&=&
- \frac{\eta}{4 \pi} \left \{ 1+ \epsilon\left[ 2\ln \left(\frac{M}{ \eta}\right)+  
2-\ln(4) \right ] \right . 
\nonumber \\
&+&  \left. \epsilon^2\left [ 4 + \frac {\pi^2}{4} + 2
\ln^2\left ( \frac{M}{ 2 \eta}\right ) + 
4 \ln\left ( \frac{M}{ 2 \eta}\right ) \right]
+{\cal O}(\epsilon^3) \right \} \;\;,
\end{eqnarray}
and its derivatives with respect to $\eta^2$:

\begin{equation}
\int_p \frac{1}{(p^2 + \eta^2)^n}=
\frac{1}{(n-1)!}\left (- \frac{d}{d \eta^2} \right )^{n-1} \int_p 
\frac{1}{p^2 + \eta^2} \;.
\label{deris}
\end{equation}

Let us now consider the three-loop contributions to $\langle \phi^2
\rangle_u$ with any number of, external and/or internal, $\delta \eta^2$
insertions. Their general form is

\begin{equation}
(\delta \eta^2)^{n-2} N \int_p \frac {\delta^c \Sigma_{a}(p)}{(p^2 +\eta^2)^n}  \;,
\end{equation}
where $c$ is defined below and $n$ determines the number of external (to
the setting sun) $\delta \eta^2$ insertions. At the same time the
insertions, internal to the setting sun, are taken into account by

\begin{equation}
\Sigma_{a}(p)= - \frac{{\cal M}(N+2)}{18}u^2 \int_{kq} \frac {1}{(k^2 +\eta^2)^m}
\frac {(\delta \eta^2)^{m+j+h-3}  }{(q^2 +\eta^2)^j}\frac {1}{[(p+k+q)^2 +
\eta^2]^h} \;,
\end{equation}
where ${\cal M}$ defines the multiplicity of equivalent internal $\delta \eta^2$ insertions.
This general contribution to  $\langle \phi^2 \rangle_u$ can be written as

\begin{eqnarray}
-(\delta \eta^2)^{n-2} N \int_p \frac {\delta^c \Sigma_{a}(p)}{(p^2 +\eta^2)^n}
&=& \delta^{n+m+j+h-3}\frac{N(N+2){\cal M}}{18}u^2 (\eta^2)^{n+m+j+h-5} 
\nonumber \\
&\times& \int_{pkq}\frac {1}{(p^2 +\eta^2)^n} \frac {1}{(k^2 +\eta^2)^m}
\frac {1}{(q^2 +\eta^2)^j}\frac {1}{[(p+k+q)^2 +\eta^2]^h} \;,
\end{eqnarray}
where $c=m+j+h-1$ labels the order of the two-loop (setting sun) self-energy term.
Now, we can merge all propagators
through the use of
standard {}Feynman
parametrization, given as usual by 

\begin{equation}
\frac {1}{a^x b^y}= \frac {\Gamma[x+y]}{\Gamma[x] \Gamma[y]} \int_0^1 
d\alpha \frac{\alpha^{x-1}(1-\alpha)^{y-1}}{[a\alpha+b(1-\alpha)]^{x+y} }
\;, \;\;\;(x,y > 0)\;.
\end{equation}
or other generalizations. One then gets

\begin{eqnarray}
(\delta \eta^2)^{n-2} N \int_p \frac {\delta^c 
\Sigma_{a}(p)}{(p^2 +\eta^2)^n}
&=&- \frac{\delta^{n+m+j+h-3}}{(4\pi)^{9/2}}\frac{N(N+2){\cal M}}{18}
\frac {u^2}{\eta}\frac{\Gamma[n+m+j+h-9/2+3\epsilon]}{\Gamma[n]
\Gamma[m]\Gamma[j]\Gamma[h]} \nonumber \\
&\times& \left(\frac{e^{\gamma_E} M^2}{\eta^2} \right)^{3\epsilon}
\int_0^1 d\alpha d\beta d\gamma \frac{ g(\alpha)g(\beta)g(\gamma)}
{[g(\alpha,\beta,\gamma)]^{n+m+j+h-9/2+3\epsilon} }\;,
\label{ssgeral}
\end{eqnarray}
where

\begin{equation}
g(\alpha)=\alpha^{j-1}(1-\alpha)^{h-1}[\alpha(1-\alpha)]^{-j-h+3/2-\epsilon}\;,
\end{equation}

\begin{equation}
g(\beta)=\beta^{m-1}(1-\beta)^{j+h-5/2+\epsilon}
[\beta(1-\beta)]^{-j-h-m+3-2\epsilon}\;,
\end{equation}

\begin{equation}
g(\gamma)=\gamma^{n-1}(1-\gamma)^{j+h+m-4+2\epsilon}\;,
\end{equation}
and

\begin{equation}
g(\alpha,\beta,\gamma)= \gamma + \frac{1-\gamma}{1-\beta}+
\frac{1-\gamma}{\beta\alpha(1-\alpha)}\;.
\end{equation}
Then, for given $n,m,j$ and $h$ one performs the expansion in $\epsilon$
keeping the poles and finite terms as usual. {}For most situations found
in the present work the integrals over the {}Feynman parameters need to be
evaluated numerically. Here we use Monte Carlo and Vegas techniques to
perform those integrations. We have taken particular care to
keep the numerical errors
less than approximately $1\%$ in our final numerical results.

One must be 
careful in carrying out the
$\epsilon$ expansion in the expression above since sometimes the
divergences can be hidden on the exponents of the {}Feynman parameters. Since
$m,n,j$ and $h$ are positive integers ($n \ge 2$, $m,j,h \ge 1$) one sees
that $g(\gamma)$ has a pole as $\epsilon \rightarrow 0$ when $j=h=m=1$
corresponding to a setting sun diagram without internal $\delta
\eta^2$ insertions. This is the only situation where one has ultraviolet 
divergences for these contributions. {}For $j=h=m=1$ the actual
divergence appears in the term $(1-\gamma)^{2\epsilon -1}$ contained in
$g(\gamma)$ and it will appear as a $1/\epsilon$ pole if one integrates

\begin{equation} 
\int_0^1 d \gamma \frac{g(\gamma)}{[g(\alpha,\beta,\gamma)]^{n+m+j+h-
9/2+3\epsilon}} \;,
\end{equation}
by parts.
The case $p=0$ follows essentially the same lines and the general result for the 
setting sun type of contribution with any internal and/or external $\delta \eta^2$ 
insertions  is

\begin{eqnarray}
(\delta \eta^2)^{n-2} N \int_p \frac {\delta^c \Sigma_{a}(0)}{(p^2 +\eta^2)^n}
&=&- \frac{\delta^{n+m+j+h-3}}{(4\pi)^3}\frac{N(N+2){\cal M}}{18}
u^2{(\eta^2)^{n-2}}\frac{\Gamma[m+j+h-3+2\epsilon]}
{\Gamma[m]\Gamma[j]\Gamma[h]} \nonumber \\
&\times& \left(\frac{e^{\gamma_E} M^2}{\eta^2} \right)^{2\epsilon}
\int_0^1 d\alpha d\beta  \frac{ g(\alpha)g(\beta)}
{[g(\alpha,\beta)]^{m+j+h-3+2\epsilon}}\int_p 
\frac {1}{(p^2 +\eta^2)^n} \;,
\label{insezero}
\end{eqnarray}
where the integral over $p$ can be readily obtained by one of the 
methods discussed in the beginning of this appendix and

\begin{equation}
g(\alpha,\beta)=\frac{1}{1-\beta}+\frac{1}{\beta\alpha(1-\alpha)}\;.
\end{equation}
Note that the gamma function in Eq. (\ref{insezero}) displays an ultraviolet
pole when $m=j=h=1$ and is finite otherwise.

The first contribution (sixth diagram of {}Fig. 3) of this
type appears at ${\cal O}(\delta^2)$ with $n=2,m=j=h=1$. 
Since there is just one graph like this, ${\cal M}=1$, and one writes

\begin{equation}
-N\int_{p}\frac{\delta^2 \Sigma_{1}(p) }{(p^2 + \eta^2)^2}= 
\delta^2 \frac{N(N+2)}{18(4\pi)^3}\frac{  u^2}{\eta}
\left [  \frac{1}{ \epsilon}+6\ln \left (\frac {M}{\eta}
\right ) - 4.93147 \right ]\;.
\label{ssp}
\end{equation}

The $p=0$ contribution is given by

\begin{equation}
N\int_p\frac {\delta^2 \Sigma_{1}(0)}{(p^2 +\eta^2)^2}=
- \delta^2 \frac{N(N+2)}{18 (4\pi)^3}\frac{u^2}{ \eta}
\left [ \frac{1}{\epsilon}  + 6 \ln \left (\frac {M}{\eta} \right ) - 
3.78069\right ]
\;.
\label{ss0}
\end{equation}
The last two equations reproduce the results found analytically by 
Braaten and Nieto
in Ref. \cite {braaten}. Note that although Eq.
(\ref{ssp}) and Eq. (\ref{ss0}) diverge, their sum is finite and
scale independent. Together, they give the contribution

\begin{equation}
-N\int_{p}\frac{\delta^2 [\Sigma_{1}(p)-\Sigma_1(0)]}{(p^2 + \eta^2)^2}= 
-{\delta^2} \frac {u^2}{\eta} \frac {N(N+2)}{18(4\pi)^3} [0.143848]\;,
\label{braaten}
\end{equation}
which is exactly the result found in our previous work, Ref. \cite
{prb}. Now, we turn to the evaluation the setting suns with insertions. The most
expedient way would be to do the replacement $\eta \rightarrow
\eta(1-\delta)^{1/2}$ and then expand the squared root to the desired
order. However, this procedure would only give the total contribution at each
order. In order to have absolute control about each single contribution
we prefer to use our general expressions Eqs. (\ref{ssgeral}) and (\ref
{insezero}). We have checked both procedures finding that they agree
to each other
within $1 \%$ which is reassuring since when obtaining the diagrams with
insertions via $\eta \rightarrow \eta(1-\delta)^{1/2}$ one has a result
which may be considered exact since this expansion starts from Eq. (\ref
{braaten}), and this result agrees with the analytical results of Refs.
\cite {prb,braaten}.
Also, at order-$\delta^4$, the general relations  given by  Eqs. (\ref{ssgeral}) and (\ref
{insezero}) have proven to be very useful in the evaluation of diagrams containing the setting sun as insertion.
{}For the eighth graph of figure 3 one has three cases similar to  
$n=m=2$ and $j=h=1$ (${\cal M}=3$). This gives

\begin{equation}
-  N\int_{p}\frac{\delta^3 \Sigma_{2}(p)}{(p^2 + \eta^2)^2}=
\delta^3 \frac{u^2}{\eta} \frac {N(N+2)}{18 (4\pi)^3}[0.188]
\;,
\end{equation}
and

\begin{equation}
N\int_{p}\frac{\delta^3 \Sigma_2(0)}{(p^2 + \eta^2)^2}=
- \delta^3 \frac{u^2}{\eta} \frac {N(N+2)}{18 (4\pi)^3}[0.249]
\;,
\end{equation}
which lead to

\begin{equation}
  N \int_{p}\frac{\delta^3 [\Sigma_{2}(p)-\Sigma_2(0)]}{(p^2 +
\eta^2)^2}=  - \delta^3 \frac {u^2}{\eta} \frac {N(N+2)}{18 (4\pi)^3}
[0.0610]\;.
\end{equation}

{}For the seventh graph of {}Fig. 3, $n=3$ and $m=h=j=1$ with ${\cal M}=1$, one has

\begin{equation}
-N\int_{p}\frac{\delta^3 \eta^2\Sigma_{1}(p)}{(p^2 + \eta^2)^3}=
\delta^3 \frac {u^2}{\eta} \frac {N(N+2)}{18(4\pi)^3}\frac{1}{32}
\left [ \frac{1}{\epsilon}
+6 \ln \left (\frac {M}{\eta} \right ) -1.968
\right ] \;,
\end{equation}
and 

\begin{equation}
N\int_{p}\frac{\delta^3 \eta^2 \Sigma_1(0)}{(p^2 + \eta^2)^3}= 
-\delta^3 \frac {u^2}{\eta} \frac {N(N+2)}{18(4\pi)^3}
\frac{1}{32} \left [ \frac{1}{\epsilon} +6 \ln \left (
\frac {M}{\eta} \right ) - 1.781 \right ] \;,
\end{equation}
which lead to 

\begin{equation}
-N\int_{p}\frac{\delta^3 \eta^2 2 [\Sigma_{1}(p)-\Sigma_1(0)]}{(p^2 + \eta^2)^3}
= -\delta^3 \frac {u^2}{\eta} \frac {N(N+2)}{18(4\pi)^3}[0.01168]  \;.
\end{equation}
where the factor of 2 on the RHS accounts for the two possibilities of
external insertions (see Eq. (\ref {expansion}).  {}For $n=4$ and
$m=h=j=1$ (tenth diagram of {}Fig. 3) one gets, with ${\cal M}=1$,

\begin{equation}
-   N\int_{p} \frac {\delta^4\eta^4\Sigma_{1}(p) }{(p^2 +\eta^2)^4}=
\delta^4 \frac {N(N+2)}{18 (4 \pi)^3 } \frac{1}{64}
\frac{u^2}{\eta}   \left [ \frac{1}{\epsilon} +6 \ln \left(
\frac {M}{\eta} \right )  - 1.20111 \right ]  \;,
\end{equation}
and

\begin{equation}
N\int_{0} \frac {\delta^4 \eta^4\Sigma_{1}(0) }{(p^2 +\eta^2)^4}=
-\delta^4 \frac {N(N+2)}{18(4 \pi)^3 } \frac {1}{64}\frac{u^2}{\eta}
\left [\frac{1}{\epsilon} +6 \ln \left (\frac {M}{\eta} \right ) -1.1408
\right ] \;,
\end{equation}
which leads to 

\begin{equation}
-   N\int_{p} \frac {3\delta^4 \eta^4  [\Sigma_{1}(p) -
\Sigma_{1}(0)]}{(p^2 +\eta^2)^4} =
-\delta^4 \frac {N(N+2)}{18(4 \pi)^3} \frac{u^2}{\eta} 
[2.8270 \times 10^{-3}]
\;,
\end{equation}
where the factor of 3 on the RHS accounts for the possibilities of
external insertions (see Eq. (\ref {expansion}).
{}There are three cases (${\cal M}=3$) similar to the case  
$n=3$, $m=2$ and $j=h=1$
displayed by the eleventh graph of {}Fig. 3. One gets

\begin{equation}
- N\int_{p}\frac {\delta^4 \eta^2\Sigma_{2}(p)}{(p^2 +\eta^2)^3}=
\delta^4 \frac {N(N+2)}{18(4\pi)^3} \frac{u^2}{\eta}  [0.0586]
\;,
\end{equation}
and

\begin{equation}
N\int_{0}\frac {\delta^4 \eta^2\Sigma_{2}(0)}{(p^2 +\eta^2)^3}= -
\delta^4 \frac {N(N+2)}{18(4\pi)^3} \frac{u^2}{\eta}  [0.0650]
\;,
\end{equation}
which lead to

\begin{equation}
-  N\int_{p}\frac { \delta^4 \eta^2  [2\Sigma_{2}(p) -
\Sigma_{2}(0)]}{(p^2 +\eta^2)^3}= - \delta^4  \frac {N(N+2)}{18(4\pi)^3}
\frac{u^2}{\eta} [7.7318\times 10^{-3}]
\;.
\end{equation}

{}The twelveth graph of {}Fig. 3 has ${\cal M}=3$, $n=2$, $m=3$ and $j=h=1$ 
leading to

\begin{equation}
-  N\int_{p} \frac {\delta^4 \Sigma_{4}(p)} {(p^2 +\eta^2)^2}=
\delta^4 \frac {N(N+2)}{18(4\pi)^3} \frac{u^2}{\eta} [0.058638]
\;,
\end{equation} 
and

\begin{equation}
 N\int_{p} \frac {\delta^4 \Sigma_{4}(0)}{(p^2 +\eta^2)^2}=-
\delta^4 \frac {N(N+2)}{18(4\pi)^3} \frac{u^2}{\eta}[0.083246]
\;,
\end{equation} 
which give

\begin{equation}
- N\int_{p} \frac {[\Sigma_{4}(p) - \Sigma_{4}(0)]}{(p^2
+\eta^2)^2}=-\delta^4 \frac {N(N+2)}{18(4\pi)^3} \frac{u^2}{\eta}   [0.02461]
\;.
\end{equation}

{}Finally, for the thirteenth graph, ${\cal M}=3$, $n=m=j=2$ and $h=1$ from which one gets

\begin{equation}
- N\int_{p}
\frac {\delta^4 \Sigma_{7}(p) }{(p^2 +\eta^2)^2}=\delta^4 \frac
{N(N+2)}{18(4\pi)^3} \frac{u^2}{\eta} [0.0234552]
\;,
\end{equation}
and

\begin{equation}
 N\int_{p}
\frac {\delta^4 \Sigma_{7}(0) }{(p^2 +\eta^2)^2}=- \delta^4 \frac
{N(N+2)}{18(4\pi)^4} \frac{u^2}{\eta}  [0.0417]
\;,
\end{equation}
which gives

\begin{equation}
-   N\int_{p}
\frac {\delta^4 [\Sigma_{7}(p) - \Sigma_{7}(0)]}{(p^2 +\eta^2)^2}=- \delta^4
\frac {N(N+2)}{18(4\pi)^3} \frac{u^2}{\eta}   [0.01825]
\;.
\end{equation}

{}Let us now consider a general four-loop contribution with any number of
internal and/or external $\delta \eta^2$ insertions. After performing
 few shifts on the integration variables one gets

\begin{eqnarray}
-(\delta \eta^2)^{n-2} N \int_p \frac {\delta^d \Sigma_{b}(p)}
{(p^2 +\eta^2)^n}
&=&- \delta^{n+m+l+h+i+j-4} u^3 N
\frac {{\cal M}\left( 16 + 10 N +  N^2 \right) }{108}
(\eta^2)^{n+m+l+h+i+j-7} \nonumber \\
&\times& \int_{pqkt} \frac{1} {(t^2 +\eta^2)^n
(q^2 +\eta^2)^m(k^2 +\eta^2)^l
[(p+q)^2 +\eta^2]^h} 
\nonumber \\
&\times& \frac{1}{[(p+k)^2 +\eta^2]^i
[(p+t)^2 +\eta^2]^j } \;,
\end{eqnarray}
where $d=m+l+h+i+j-2$ labels the order of the three-loop  self-energy
term. Then, proceeding as in the three loop case one finds

\begin{eqnarray}
-(\delta \eta^2)^{n-2} N \int_p \frac {\delta^d \Sigma_{b}(p)}
{(p^2 +\eta^2)^n}
&=&-N \frac{ \delta^{n+m+l+h+i+j-4}}{(4\pi)^6} \frac {u^3}{\eta^2} 
\frac {{\cal M} \left( 16 + 10 N +  N^2 \right) }{108}
\left ( \frac {M^2 \exp \gamma_E}{\eta^2} \right )^{4\epsilon}
\nonumber \\
&\times& \frac {\Gamma(n+m+l+h+i+j -6 +4 \epsilon)}{\Gamma(n)
\Gamma(m)\Gamma(l)\Gamma(h)\Gamma(i)\Gamma(j)} \nonumber \\
&\times& \int_0^1 d\alpha d\beta d\gamma d\theta d\phi 
\frac{f(\alpha)f(\beta)
f(\gamma)f(\theta)f(\phi)}
{[f(\alpha,\beta,\gamma,\theta,\phi)]^{l+i+m+h+n+j-6+4\epsilon}} \;\;,
\end{eqnarray} 
where

\begin{equation}
f(\alpha) = \alpha^{1/2-i-\epsilon}(1-\alpha)^{1/2-l-\epsilon}\;,
\label{fa}
\end{equation}

\begin{equation}
f(\beta) = \beta^{1/2-h-\epsilon}(1-\beta)^{1/2-m-\epsilon}\;,
\label{fb}
\end{equation}

\begin{equation}
f(\gamma) = \gamma^{1/2-j-\epsilon}(1-\gamma)^{1/2-n-\epsilon}\;,
\end{equation}

\begin{equation}
f(\theta) = \theta^{m+h-5/2+\epsilon}(1-\theta)^{n+j-5/2+\epsilon}\;,
\end{equation}

\begin{equation}
f(\phi) = \phi^{l+i-5/2+\epsilon}(1-\phi)^{n+j+m+h-4+2\epsilon}\;,
\end{equation}
and

\begin{equation} f(\alpha,\beta,\gamma,\theta,\phi)=
\frac{\phi}{\alpha(1-\alpha)} + \frac{\theta(1-\phi)}{\beta(1-\beta)} +
\frac{(1-\theta)(1-\phi)}{\gamma(1-\gamma)} \; . 
\end{equation} 
As far  as
renormalization is concerned one should note that those type of four-loop 
contributions are always finite. The four-loop contribution whose
self-energy has zero external momentum reads

\begin{equation}
(\delta \eta^2)^{n-2} N \int_p \frac {\delta^d \Sigma_{b}(0)}{(p^2 +\eta^2)^n} \;,
\end{equation}
where

\begin{eqnarray}
\Sigma_{b}(0)&=& N \delta^3 u^3
\frac{{\cal M}\left( 16 + 10 N +  N^2 \right) }{108}
(\delta \eta^2)^{n-2} 
(\delta \eta^2)^{m+l+h+i+j-5} \nonumber \\
&\times& \int_{qkt} \frac{1} {(q^2 +\eta^2)^j(k^2 +\eta^2)^l(t^2 +\eta^2)^h
[(q+k)^2 +\eta^2]^i[(q+t)^2 +\eta^2]^m } \;.
\end{eqnarray}
Proceeding as above one gets

\begin{eqnarray}
(\delta \eta^2)^{n-2} N \int_p \frac {\delta^d \Sigma_{b}(0)}
{(p^2 +\eta^2)^n}&=&
\delta^{n+m+l+h+i+j-4} N \frac{u^3(\eta^2)^{n-5/2}}{(4\pi)^{9/2}}
\frac{{\cal M}\left( 16 + 10 N +  N^2 \right) }{108}
\nonumber \\
&\times& \left ( \frac {M^2 \exp \gamma_E}{\eta^2} \right )^{3\epsilon} 
\frac {\Gamma(m+l+h+i+j -9/2 +3 \epsilon)}{\Gamma(m)\Gamma(l)\Gamma(h)
\Gamma(i)\Gamma(j)} 
\nonumber \\
&\times&  \int_0^1 d\alpha d\beta d\gamma d\theta  
\frac{\gamma f(\alpha)f(\beta)
f(\gamma,\theta)}{[f(\alpha,\beta,\gamma,\theta)]^{l+i+m+h+j-9/2+3\epsilon}}
\nonumber \\
&\times& \int_p \frac {1}{(p^2 +\eta^2)^n}\;\;,
\end{eqnarray}
where $f(\alpha)$ and $f(\beta)$ are given by Eqs. ({\ref {fa}) and ({\ref {fb}). 
Also, one has

\begin{equation}
f(\gamma,\theta)=(1-\gamma)^{j-1}[\gamma(1-\theta)]^{l+i+\epsilon-5/2} 
(\gamma \theta)^{h+m-5/2+\epsilon} \;,
\end{equation}
and

\begin{equation}
f(\alpha,\beta,\gamma,\theta)= (1-\gamma) + 
\frac {\gamma(1-\theta)}{\alpha(1-\alpha)} + 
\frac{\gamma\theta}{\beta(1-\beta)}\; .
\end{equation}
The first four-loop contribution of this type appears at
order-$\delta^3$ and is displayed by the ninth graph of {}Fig. 3 and it
has $n=2$, $m=l=h=i=j=1$ and ${\cal M}=1$.  The contributions to $\langle \phi^2
\rangle_u$ are given by

\begin{equation}
-\delta^3N\int_{p}\frac {\Sigma_{3}(p) }{(p^2 +\eta^2)^2} = 
- \delta^3 \frac {u^3}{\eta^2}
\frac{N}{(4\pi)^6}
\frac{\left( 16 + 10 N +  N^2 \right) }{108} [32.4388]\;,
\end{equation}
and

\begin{equation}
\delta^3N\int_{p}\frac {\Sigma_{3}(0) }{(p^2 +\eta^2)^2} = 
\delta^3 \frac {u^3}{\eta^2} \frac{N}{(4\pi)^6}
\frac{\left( 16 + 10 N +  N^2 \right) }{108} [40.538]\;,
\end{equation}
which lead to

\begin{equation}
-\delta^3N\int_{p}\frac {[\Sigma_{3}(p) - 
\Sigma_{3}(0)]}{(p^2 +\eta^2)^2}=\delta^3 \frac {u^3}{\eta^2} 
\frac{N}{(4\pi)^6}
\frac{\left( 16 + 10 N +  N^2 \right) }{108} [8.09927]\;.
\end{equation}
As for the three-loop case one could use the equation above to obtain a
series expansion which would give the total contribution of graphs with
insertions to any order in $\delta$. However, we prefer to perform the
individual evaluation of each contribution in order
to achieve more control over the series expansion. At order-$\delta^4$
the first contribution is displayed by the fourteenth graph with ${\cal
M}=1$, $n=3$ and $m=l=h=i=j=1$. One then obtains

\begin{equation}
- \delta^4 N\int_{p} \frac {\eta^2\Sigma_{3}(p) }
{(p^2 +\eta^2)^3}=- \delta^4 \frac {u^3}{\eta^2} \frac{N}{(4\pi)^6}
\frac{\left( 16 + 10 N +  N^2 \right) }{108} [9.70448]\;,
\end{equation} 
and

\begin{equation}
\delta^4 N\int_{p} \frac {\eta^2\Sigma_{3}(0) }{(p^2 +\eta^2)^3}=
\delta^4 \frac {u^3}{\eta^2} \frac{N}{(4\pi)^6}
\frac{\left( 16 + 10 N +  N^2 \right) }{108} [10.1344]\;,
\end{equation}
which lead to

\begin{equation}
-  \delta^4 N\int_{p} \frac {2\eta^2 [\Sigma_{3}(p) -
\Sigma_{3}(0)]}{(p^2 +\eta^2)^3}=
\delta^4 \frac {u^3}{\eta^2} \frac{N}{(4\pi)^6}
\frac{\left( 16 + 10 N +  N^2 \right) }{108} [0.85984]
\;,
\end{equation}
where, once more, the factor of 2 accounts for the two possibilities of 
internal insertions in accordance with Eq. ({\ref {expansion}).

Next, let us consider the case illustrated by the fifteenth graph 
of {}Fig. 3, which has ${\cal M}=1$, $n=j=2$ and $h=i=l=m=1$.
After evaluating the integrals one gets

\begin{equation}
-\delta^4 N\int_{p}\frac {\Sigma_{10}(p) }{(p^2 +\eta^2)^2}=
-\delta^4 \frac {u^3}{\eta^2}
\frac{N}{(4\pi)^6}
\frac{\left( 16 + 10 N +  N^2 \right) }{108} [3.18221]\;,
\end{equation}
and

\begin{equation}
\delta^4 N\int_{p}\frac {\Sigma_{10}(0) }{(p^2 +\eta^2)^2}=
\delta^4 \frac {u^3}{\eta^2} \frac{N}{(4\pi)^6}
\frac{\left( 16 + 10 N +  N^2 \right) }{108} [5.120]\;,
\end{equation}
which lead to

\begin{equation}
-\delta^4 N\int_{p}\frac {[\Sigma_{10}(p) - \Sigma_{10}(0)]}
{(p^2 +\eta^2)^2}=
\delta^4 \frac {u^3}{\eta^2} \frac{N}{(4\pi)^6}
\frac{\left( 16 + 10 N +  N^2 \right) }{108} [1.9784]\;.
\end{equation}

The remaining  four-loop contributions to this order are evaluated 
using the case displayed by the sixteenth diagram of {}Fig. 3 which has
${\cal M}=4$, $n=m=2$, $l=h=i=j=1$ and whose result is given by

\begin{equation}
- N\int_{p}\frac {\delta^4 \Sigma_{5}(p) }{(p^2 +\eta^2)^2}=
- \delta^4 \frac {u^3}{\eta^2} \frac{N}{(4\pi)^6}
\frac{\left( 16 + 10 N +  N^2 \right) }{108} [9.85072]\;,
\end{equation}
and

\begin{equation}
N\int_{p}\frac {\delta^4 \Sigma_{5}(0) }{(p^2 +\eta^2)^2}=
\delta^4 \frac {u^3}{\eta^2} \frac{N}{(4\pi)^6}
\frac{\left( 16 + 10 N +  N^2 \right) }{108} [15.15548]\;,
\end{equation}  
which lead to

\begin{equation}
-  N\int_{p}\frac {\delta^4[\Sigma_{5}(p) - \Sigma_{5}(0)]}{(p^2
+\eta^2)^2}= \delta^4  \frac {u^3}{\eta^2} \frac{N}{(4\pi)^6}
\frac{\left( 16 + 10 N +  N^2 \right) }{108} [5.30476]\;.
\end{equation}

Let us now consider the five-loop contributions.
The first one is given by  the seventeenth graph of {}Fig. 3,

\begin{equation}
\delta^4 N\int_{p}\frac {[\Sigma_{1}(p) - \Sigma_{1}(0)]^2}
{(p^2 +\eta^2)^3}= -\delta^4
\frac {u^4}{\eta^3} \frac {N(N+2)^2}{ (18)^2 (4\pi)^7}[0.87339]\;,
\end{equation}
where the individual contributions are given by three terms starting with

\begin{eqnarray}
\delta^4 N\int_{p}\frac {[\Sigma_1(p)]^2}{(p^2 +\eta^2)^3}&=&  \delta^4
\frac {u^4}{\eta^3} \frac {N(N+2)^2\Gamma(3/2+5\epsilon)}{(18)^2 
\Gamma(3) (4\pi)^{15/2}}\left ( \frac {M^2 \exp \gamma_E}{\eta^2} 
\right )^{5\epsilon}
\nonumber \\
&\times& \int_0^1 d\alpha d\beta d\gamma d\theta d\phi d\chi 
\frac{h(\alpha)h(\beta)
h(\gamma)h(\theta)h(\phi)h(\chi)}{[
h(\alpha,\beta,\gamma,\theta,\phi,\chi)]^{3/2+5\epsilon}} 
\nonumber\\
&\times& \left \{ \frac {1}{\epsilon^2} \left [ 1 - \frac {3}{4} 
\frac {\phi h(\alpha,\beta,\gamma,\theta,\chi)}
{h(\alpha,\beta,\gamma,\theta,\phi,\chi)} - \frac {9}{8} 
\frac {\phi \gamma[1- h(\alpha,\beta)]}
{h(\alpha,\beta,\gamma,\theta,\phi,\chi)}\right . 
\right . \nonumber \\
&+& \left . \frac {15}{16} \frac {\phi^2 \gamma[1- h(\alpha,\beta)]
h(\alpha,\beta,\gamma,\theta,\chi)}{[h(\alpha,\beta,\gamma,\theta,\phi,\chi)]^2} 
\right ]  \nonumber \\
&+& \frac {1}{\epsilon} \left [ 1 - \frac {5}{2} \frac {\phi 
h(\alpha,\beta,\gamma,\theta,\chi)}{h(\alpha,\beta,\gamma,\theta,\phi,\chi)}
- \frac {18}{4} \frac {\phi \gamma [1- h(\alpha,\beta)]}{
h(\alpha,\beta,\gamma,\theta,\phi,\chi)} \right . 
\nonumber \\
&+& \left . \frac {40}{8} \frac {\phi^2 \gamma [1-
h(\alpha,\beta)]h(\alpha,\beta,\gamma,\theta,\chi)}
{[h(\alpha,\beta,\gamma,\theta,\phi,\chi)]^2} \right ] 
\nonumber \\
&-& \left . \frac {5}{2} \frac {\phi \gamma[1- h(\alpha,\beta)]}
{h(\alpha,\beta,\gamma,\theta,\phi,\chi)}
+ \frac {25}{4} \frac {\phi^2 \gamma [1-h(\alpha,\beta)]
h(\alpha,\beta,\gamma,\theta,\chi)}
{[h(\alpha,\beta,\gamma,\theta,\phi,\chi)]^2} \right \} \;,
\end{eqnarray}
where

\begin{equation}
h(\alpha) = [\alpha(1-\alpha)]^{-1/2-\epsilon}\;,
\end{equation}

\begin{equation}
h(\beta) = (1-\beta)^{-1/2+\epsilon}[\beta(1-\beta)]^{-2\epsilon}\;,
\end{equation}

\begin{equation}
h(\gamma) = \gamma(1-\gamma)^{2\epsilon}\;,
\end{equation}

\begin{equation}
h(\theta) = [\theta(1-\theta)]^{-1/2-\epsilon}\;,
\end{equation}

\begin{equation}
h(\phi) = \phi^{1+2\epsilon}(1-\phi)^{2\epsilon}\;,
\end{equation}

\begin{equation}
h(\chi) = (1-\chi)^{-1/2+\epsilon}[\chi(1-\chi)]^{-2\epsilon}\;,
\end{equation}

\begin{equation}
h(\alpha,\beta) = \frac{1}{1-\beta} + \frac {1}{\beta\alpha(1-\alpha)}\;,
\end{equation}

\begin{equation}
h(\theta,\chi) = \frac{1}{1-\chi} + \frac {1}{\chi\theta(1-\theta)}   \,,
\end{equation}

\begin{equation}
h(\alpha,\beta,\gamma,\theta,\chi)= \gamma  - h(\theta,\chi) +
(1-\gamma)h(\alpha,\beta) \;,
\end{equation}
and 

\begin{equation}
h(\alpha,\beta,\gamma,\theta,\phi,\chi)= \gamma \phi + (1-\phi)h(\theta,\chi) +
\phi(1-\gamma)h(\alpha,\beta)\;.
\end{equation}
After performing the expansion in $\epsilon$ and integrating numerically one obtains

\begin{eqnarray}
\delta^4 N\int_{p}\frac {[\Sigma_{1}(p)]^2}{(p^2 +\eta^2)^3}&=& \delta^4 
\frac {u^4}{\eta^3} \frac {N(N+2)^2}{ 1296 (8\pi)^5}
\left \{ \frac {1}{\epsilon^2} \right . 
\nonumber \\
&+& \left . \frac{1}{\epsilon} \left [ 10 \ln \left ( \frac {M}{\eta} \right ) - 
4.419\right ]
+ 50 \ln^2 \left ( \frac {M}{\eta} \right )
- (44.19)\ln \left ( \frac {M}{\eta} \right )+ 
20.0158 \right \}
\;.
\end{eqnarray}

Now,  expanding Eq. (\ref{ssgeral}), with $n=3,m=j=h=1$ to order 
$\epsilon$ and   considering

\begin{equation}
\Sigma_1(0)= - \delta^2  \frac {u^2}{(8 \pi)^2} \frac {(N+2)}{18} 
\left \{ \frac {1}{\epsilon} + 4 \ln \left ( \frac {M}{\eta} \right )
- 2.3911
+ \epsilon \left [ 8 \ln^2\left ( \frac {M}{\eta} \right ) + \frac{\pi^2}{3} - 
9.5644 \ln \left ( \frac {M}{\eta} \right )
+ 4.3127 \right ] \right \} \;,
\label{ssunzero}
\end{equation}
one gets

\begin{eqnarray}
-2 \delta^4 N\int_{p}\frac {[\Sigma_{1}(p) \times \Sigma_{1}(0)]}
{(p^2 +\eta^2)^3}&=&  - \delta^4 \frac {u^4}{\eta^3}
 \frac {N(N+2)^2} {648 (8\pi)^5} \left \{ \frac {1}{\epsilon^2} 
\right . \nonumber \\
&+&\left . \frac{1}{\epsilon} \left [ 10 \ln 
\left ( \frac {M}{\eta} \right ) - 4.3067 \right ]
+ 50 \ln^2 \left ( \frac {M}{\eta} \right )
- (42.937)\ln \left ( \frac {M}{\eta} \right ) +
18.9756 \right \} \;.
\end{eqnarray}
The final contribution to this diagram is obtained by considering 
Eq. (\ref{intsimples}), with $n=3$, expanded to order $\epsilon^2$
and by taking the square of Eq. (\ref{ssunzero}) which leads to

\begin{eqnarray}
\delta^4 N\int_{p}\frac {[\Sigma_{1}(0)]^2}{(p^2 +\eta^2)^3}&=& 
\delta^4 \frac {u^4}{\eta^3} \frac {N(N+2)^2}{ 1296 (8\pi)^5}
\left \{ \frac {1}{\epsilon^2} \right . \nonumber \\
&+& \left .\frac{1}{\epsilon} \left [ 10 \ln \left ( 
\frac {M}{\eta} \right ) - 4.1684 \right ]
+ 50 \ln^2 \left ( \frac {M}{\eta} \right )
- (41.684)\ln \left ( \frac {M}{\eta} \right ) + 18.6434 \right \}
\;.
\end{eqnarray}

Next, let us consider the eighteenth graph of {}Fig. 3 whose 
contribution comes from

\begin{eqnarray}
-\delta^4 N\int_{p} \frac {\Sigma_{6}(p) }{(p^2 +\eta^2)^2}&=&
 \delta^4 \frac {u^4}{\eta^3}\frac{N}
{(4\pi)^{15/2}} 
\frac{\left( 40 + 32 N +8 N^2 + N^3 \right) }{648}
\Gamma(3/2+5\epsilon)\left ( \frac {M^2 \exp \gamma_E}{\eta^2} 
\right )^{5\epsilon}\nonumber \\
&\times& \int_0^1 d\alpha d\beta d\gamma d\theta d\phi d\chi d\zeta 
\frac{y(\alpha)y(\beta)
y(\gamma)y(\theta)y(\phi)y(\chi)y(\zeta)}{[y(\alpha,\beta,\gamma,
\theta,\phi,\chi,\zeta)]^{3/2+5\epsilon}}
\;.
\end{eqnarray}
where

\begin{equation}
y(\alpha) = \alpha [\alpha(1-\alpha)]^{-3/2-\epsilon}\;,
\end{equation}

\begin{equation}
y(\beta) = [\beta(1-\beta)]^{-1/2-\epsilon}\;,
\end{equation}

\begin{equation}
y(\gamma) = [\gamma(1-\gamma)]^{-1/2-\epsilon}\;,
\end{equation}

\begin{equation}
y(\theta) = [\theta(1-\theta)]^{-1/2-\epsilon}\;,
\end{equation}

\begin{equation}
y(\phi) = \phi^{1/2+\epsilon}(1-\phi)^{-1/2+\epsilon}\;,
\end{equation}

\begin{equation}
y(\chi) = [\chi(1-\chi)]^{-1/2+\epsilon}\;,
\end{equation}

\begin{equation}
y(\zeta)= \zeta^{1+2\epsilon}(1-\zeta)^{2\epsilon} \;,
\end{equation}
and
\begin{equation}
y(\alpha,\beta,\gamma,\theta,\phi,\chi,\zeta)= \frac{\zeta(1-\phi)}
{\beta(1-\beta)} +\frac{\phi \zeta}{\alpha(1-\alpha)} +
\frac {(1-\chi)(1-\zeta)}{\theta(1-\theta)} + \frac{\chi 
(1- \zeta)}{\gamma(1-\gamma)} \;.
\end{equation}
This contribution is finite and yields

\begin{eqnarray}
-\delta^4 N\int_{p} \frac {\Sigma_{6}(p) }{(p^2 +\eta^2)^2}&=&
\delta^4 \frac {u^4}{\eta^3}\frac{N}{(4\pi)^{7}}
\frac{\left( 40 + 32 N +8 N^2 + N^3 \right) }{648}
[7.05619233] \;.
\end{eqnarray}

The $p=0$ case is given by

\begin{eqnarray}
\delta^4 N\int_{p} \frac {\Sigma_{6}(0) }{(p^2 +\eta^2)^2}&=&
- \delta^4 \frac {u^4}{\eta^2}\frac{N}
{(4\pi)^{6}}
\frac{\left( 40 + 32 N +8 N^2 + N^3 \right) }{648}
\Gamma(1+4\epsilon)\left ( \frac {M^2 \exp 
\gamma_E}{\eta^2} \right )^{4\epsilon}
\nonumber \\
&\times& \int_0^1 d\alpha d\beta d\gamma d\theta d\phi d\chi 
\frac{y{\prime}(\alpha)y(\beta)
y(\gamma)y{\prime}(\theta)y{\prime}(\phi)y(\chi)}{[y(\alpha,
\beta,\gamma,\theta,\phi,\chi)]^{1+4\epsilon}} \int_{p} 
\frac {1}{(p^2 +\eta^2)^2} \;,
\end{eqnarray}
where

\begin{equation}
y{\prime}(\alpha)=[\alpha(1-\alpha)]^{-1/2-\epsilon}  \;,
\end{equation}

\begin{equation}
y{\prime}(\theta)=(1-\theta)^{-1/2+\epsilon}   \;,
\end{equation}

\begin{equation}
y{\prime}(\phi)= \phi^{1/2+\epsilon}(1-\phi)^{2\epsilon }  \;,
\end{equation}

\begin{equation}
y(\alpha,\beta,\gamma,\theta,\phi,\chi)= \phi \theta + 
\frac{\phi(1-\theta)}{\alpha(1-\alpha)} + 
\frac{\chi(1-\phi)}{\beta(1-\beta)} + \frac {(1-\chi)(1-\phi)}{\gamma(1-\gamma)}
\;.
\end{equation}
Integrating one obtains

\begin{eqnarray}
\delta^4 N\int_{p} \frac {\Sigma_{6}(0) }{(p^2 +\eta^2)^2}&=&
-\delta^4 \frac {u^4}{\eta^3}\frac{N}{(4\pi)^{7}}
\frac{\left( 40 + 32 N +8 N^2 + N^3 \right) }{648}
[10.21524]
\;.
\end{eqnarray}

Together these contributions yield

\begin{equation}
-\delta^4 N\int_{p} \frac {[\Sigma_{6}(p) - 
\Sigma_{6}(0)]}{(p^2 +\eta^2)^2}=- \delta^4 \frac {u^4}{\eta^3}
\frac{N}{(4\pi)^7}
\frac{\left( 40 + 32 N +8 N^2 + N^3 \right) }{648} [3.15904767]
\;.
\end{equation}

The nineteenth contribution of {}Fig. 3 is given by

\begin{equation}
-N\int_{p} \frac {[\Sigma_{8}(p) - \Sigma_{8}(0)]}{(p^2 +\eta^2)^2} =-
\delta^4 \frac {u^4}{\eta^3}\frac{N}{(4\pi)^{7}}
\frac{\left( 44 + 32 N + 5 N^2  \right) }{324}
[1.70959]  \;.
\end{equation}
The first contribution to this result follows from

\begin{eqnarray}
-\delta^4 N\int_{p} \frac {\Sigma_{8}(p) }{(p^2 +\eta^2)^2}&=&
\delta^4 \frac {u^4}{\eta^3}
\frac{N}{(4\pi)^{15/2}}
\frac{\left( 44 + 32 N + 5N^2  \right) }{324}
\Gamma(3/2+5\epsilon)\left (
\frac {M^2 \exp \gamma_E}{\eta^2} \right )^{5\epsilon}
\nonumber \\
&\times& \int_0^1 d\alpha d\beta d\gamma d\theta d\phi 
d\chi d\xi \frac{k(\alpha)k(\beta)
k(\theta)k(\phi)k(\chi) k(\gamma) k(\xi) k(\phi,\theta)
\Lambda_8^{-3/2-3 \epsilon}}
{[1- \gamma+\Xi_8\gamma]^{3/2+5\epsilon}}
\;,
\label{olhop}
\end{eqnarray}
where

\begin{equation}
k(\alpha)=[\alpha(1-\alpha)]^{-1/2-\epsilon}  \;,
\end{equation}

\begin{equation}
k(\beta)=[\beta(1-\beta)]^{-1-4\epsilon}\beta^{1/2+3 \epsilon}  \;,
\end{equation}

\begin{equation}
k(\phi)= \phi^{-3/2- \epsilon} \;,
\end{equation}

\begin{equation}
k(\theta)=(1-\theta)^{-1/2-\epsilon}   \;,
\end{equation}

\begin{equation}
k(\chi)= (1-\chi)^{1+2\epsilon }  \;,
\end{equation}

\begin{equation}
k(\gamma)= (1-\gamma) \gamma^{4\epsilon }  \;,
\end{equation}

\begin{equation}
k(\xi)= \xi^{2 \epsilon}   \;,
\end{equation}

\begin{equation}
k(\phi,\theta) =\left[1-\phi (1-\theta)\right]^{-1-2\epsilon} \;,
\end{equation}

\begin{equation}
\Lambda_8= \chi +\frac{\theta^2 \xi (1-\chi)}{[1-\phi (1-\theta)]^2}-
\left[ \chi + \frac{\theta \xi (1-\chi)}{1-\phi (1-\theta)}\right]^2
- \frac{\theta^2 \xi(1-\chi)}{1-\phi(1-\theta)}+\frac{\xi(1-\chi)\theta(1-\theta)}
{\phi(1-\theta)[1-\phi(1-\theta)]}\;,
\end{equation}
and
\begin{equation}
\Xi_8 = \frac{1-\beta+ \beta +
\beta\;\Phi_8/\Lambda_8}{\beta(1-\beta)}\;,
\end{equation}
with

\begin{equation}
\Phi_8 = 1-\xi(1-\chi) +\xi(1-\chi)\frac{1-\phi(1-\theta)+
\frac{\phi(1-\theta)}{\alpha(1-\alpha)}}{\phi(1-\theta)
[1-\phi(1-\theta)]}\;.
\end{equation}

Integrating one obtains

\begin{eqnarray}
-\delta^4 N\int_{p} \frac {\Sigma_{8}(p) }{(p^2 +\eta^2)^2}&=&
\delta^4 \frac {u^4}{\eta^3}
\frac{N}{(4\pi)^{7}}
\frac{\left( 44 + 32 N + 5N^2  \right) }{324} [4.31098]
\end{eqnarray}

The $p=0$ case is given by

\begin{eqnarray}
\delta^4 N\int_{p} \frac {\Sigma_{8}(0) }{(p^2 +\eta^2)^2}&=&
- \delta^4 \frac {u^4}{\eta^3}\frac{N}
{(4\pi)^{15/2}}
\frac{\left( 44 + 32 N + 5N^2  \right) }{324}
\Gamma(1+4\epsilon)  \Gamma(1/2+\epsilon) \left (
\frac {M^2 \exp \gamma_E}{\eta^2} \right )^{4\epsilon}
\nonumber \\
&\times& \int_0^1 d\alpha d\beta d\xi d\theta d\phi d\chi 
\frac{k(\alpha)k(\beta)
k(\xi)k(\theta)k(\phi)k(\chi) k(\phi,\theta)\Lambda_8^{-3/2-3\epsilon}
}{\Xi_8^{1+4\epsilon}} 
\;,
\label{olho0}
\end{eqnarray}
with the same notation as used in Eq. (\ref{olhop}).
Integrating Eq. (\ref{olho0}) one obtains

\begin{eqnarray}
\delta^4 N\int_{p} \frac {\Sigma_{8}(0) }{(p^2 +\eta^2)^2}&=&
-\delta^4 \frac {u^4}{\eta^3}
\frac{N}{(4\pi)^{7}}
\frac{\left( 44 + 32 N + 5N^2  \right) }{324} [6.02057]
\;.
\end{eqnarray}

We have another five-loop contribution given by  the twentieth graph 
of {}Fig. 3 whose contribution is

\begin{equation}
-N\int_{p} \frac {[\Sigma_{9}(p) - \Sigma_{9}(0)]}{(p^2 +\eta^2)^2}=
- \delta^4
\frac {u^4}{\eta^3} \frac {{\cal M} N(N+2)^2}{(18)^2(4\pi)^7}[1.4803]
\;,
\label{ciclope}
\end{equation}
where ${\cal M}=3$ accounts for the three possible ways of inserting one 
setting sun within another graph of the same type.
The first term on the LHS of Eq. (\ref {ciclope}) is
\begin{eqnarray}
\lefteqn{ -\delta^4 N\int_{p} \frac {\Sigma_{9}(p)}{(p^2 +\eta^2)^2} = -
\delta^4 \frac {N(N+2)}{18}u^2 
\int_{pkq} \frac {1}{(p^2 +\eta^2)^2}
\frac {1}{(k^2 +\eta^2)}\frac {[\Sigma_1 (q) - 
\Sigma_1(0)]}  {(q^2 +\eta^2)^2}\frac {1}{[(p+k+q)^2 +\eta^2]} }
\nonumber \\
&=&  \delta^4 \frac {u^4}{\eta^3} \frac {N(N+2)^2\Gamma(3/2+5\epsilon)}{(18)^2  
(4\pi)^{15/2}}\left ( \frac {M^2 \exp \gamma_E}{\eta^2} \right )^{5\epsilon}
\int_0^1 d\alpha d\beta d\gamma d\theta d\phi 
d\chi \frac{x(\alpha)x(\beta)
x(\gamma)x(\theta)x(\phi)x(\chi)}{[x(\alpha,\beta,\gamma,
\theta,\phi,\chi)]^{3/2+5\epsilon}} 
\nonumber\\
& &\times \left\{ \frac{1}{2\epsilon}\left [ 1 - 
\frac {3}{2}\frac{ x^{\prime}(\alpha,\beta,\gamma,\theta,\phi,\chi)}
{x(\alpha,\beta,\gamma,\theta,\phi,\chi)}\right  ] - \frac {5}{2}  
\frac{ x^{\prime}(\alpha,\beta,\gamma,\theta,\phi,\chi)}
{x(\alpha,\beta,\gamma,\theta,\phi,\chi)} \right \} 
\nonumber \\
&& -  \delta^4 \frac {u^4}{\eta^3} \frac {N(N+2)^2}{(18)^2}
\frac {[\pi \times 10^{-5}]}{(8\pi)^2} \left [ \frac {1}{\epsilon} + 
10 \ln \left ( \frac{M}{\eta} \right ) -5.9258 \right ]
\;.
\end{eqnarray}

Note that $\Sigma_1(q)$ and $\Sigma_1(0)$ have, except for the labeling
of momenta, the same form as Eqs. (\ref{ssgeral}) and (\ref{insezero}) 
with $m=j=h=1$. The $x$
functions are given by

\begin{equation}
x(\alpha) = [\alpha(1-\alpha)]^{-1/2-\epsilon}\;,
\end{equation}

\begin{equation}
x(\beta) = (1-\beta)^{-1/2+\epsilon}[\beta(1-\beta)]^{-2\epsilon}\;,
\end{equation}

\begin{equation}
x(\gamma) = (1-\gamma)^{2\epsilon}\;,
\end{equation}

\begin{equation}
x(\theta) = \theta^{1+2\epsilon}[\theta(1-\theta)]^{-3/2-3\epsilon}\;,
\end{equation}

\begin{equation}
x(\phi) = (1-\phi)^{1/2 +3\epsilon}[\phi(1-\phi)]^{-1-4\epsilon}\;,
\end{equation}

\begin{equation}
x(\chi) = \chi(1-\chi)^{4\epsilon}\;,
\end{equation}

\begin{equation}
x(\alpha,\beta) = \frac{1}{1-\beta} + \frac {1}{\beta\alpha(1-\alpha)}
\;,
\end{equation}

\begin{equation}
x(\alpha,\beta,\gamma,\theta,\phi,\chi)= \chi + x(\alpha,\beta) 
\frac{(1-\gamma)(1-\chi)}{\phi(1-\theta)}
+ \frac{\gamma(1-\chi)}{\phi(1-\theta)} + \frac{(1-\chi)}{(1-\phi)}
+ \frac{(1-\chi)}{\theta \phi} 
\;,
\end{equation}
and
\begin{equation}
x^{\prime}(\alpha,\beta,\gamma,\theta,\phi,\chi)=
\frac {\gamma(1-\chi)}{\phi(1-\theta)}[1-x(\alpha,\beta)]
\;.
\end{equation}
Then after integrating over the Feynman parameters and expanding in 
$\epsilon$ one gets

\begin{eqnarray}
-\delta^4 N\int_{p} \frac {\Sigma_{9}(p)}{(p^2 +\eta^2)^2}&=&
\delta^4 \frac {u^4}{\eta^3} \frac {N(N+2)^2}{(18)^2} 
\frac {u^4}{\eta^3} \frac{[\pi\times 10^{-5}]}{(8\pi)^2}
\left [ \frac{1}{\epsilon}
+  10 \ln \left ( \frac{M}{\eta} \right ) -6.83485 \right ] 
\nonumber \\
&-&  \delta^4 \frac {u^4}{\eta^3} \frac {N(N+2)^2}{(18)^2} 
\frac {u^4}{\eta^3} \frac{[\pi\times 10^{-5}]}{(8\pi)^2}
\left [ \frac{1}{\epsilon}
+  10 \ln \left ( \frac{M}{\eta} \right ) - 5.9258 \right ]
\;,
\end{eqnarray}
which gives the finite, scale independent result

\begin{equation}
-\delta^4 N\int_{p} \frac {\Sigma_{9}(p)}{(p^2 +\eta^2)^2}=
-\delta^4 \frac {u^4}{\eta^3} \frac {N(N+2)^2}{(18)^2} 
\frac {u^4}{\eta^3} \frac{[\pi\times 10^{-5}]}{(8\pi)^2}[0.90905]
\;.
\end{equation}

The other contribution is given by

\begin{eqnarray}
\lefteqn{ \delta^4 N\int_{p} \frac {\Sigma_{9}(0)}{(p^2 +\eta^2)^2} =
\delta^4 \frac {N(N+2)}{18}u^2 
\int_{pkq} \frac {1}{(p^2 +\eta^2)^2}\frac {1}{(k^2 +\eta^2)}
\frac {[\Sigma_1 (q) - \Sigma_1(0)]}  {(q^2 +\eta^2)^2}
\frac{1}{[(k+q)^2 +\eta^2]} }
\nonumber \\
&& =  - \delta^4 \frac {u^4}{\eta^2} \frac {N(N+2)^2
\Gamma(1+4\epsilon)}{(18)^2  (4\pi)^{6}}\left ( 
\frac {M^2 \exp \gamma_E}{\eta^2} \right )^{4\epsilon}
\int_0^1 d\alpha d\beta d\gamma d\theta d\phi  
\frac{x(\alpha)x(\beta)
x(\gamma)x(\theta)x(\phi)}{[x(\alpha,\beta,\gamma,\theta,
\phi)]^{1+4\epsilon}} \nonumber\\
&& \times \left\{ \frac{1}{2\epsilon}\left [ 1 - 
\frac{ x^{\prime}(\alpha,\beta,\gamma,\theta,\phi)}
{x(\alpha,\beta,\gamma,\theta,\phi)}\right  ] - 2  
\frac{ x^{\prime}(\alpha,\beta,\gamma,\theta,\phi)}
{x(\alpha,\beta,\gamma,\theta,\phi)} \right \} 
\int_p \frac {1}{(p^2 +\eta^2)^2}  
\nonumber \\
&& +  \delta^4 \frac {u^4}{\eta^3} \frac {N(N+2)^2}{(18)^2} 
\frac {[4.1906 \times 10^{-5}]}{(8\pi)^2} \left [ \frac {1}{\epsilon}
+ 10 \ln \left ( \frac{M}{\eta} \right ) - 6.17383 \right ] \;,
\end{eqnarray}
where

\begin{equation}
x(\alpha,\beta,\gamma,\theta,\phi)=x(\alpha,\beta) 
\frac{(1-\gamma)}{\phi(1-\theta)}+\frac{\gamma}{\phi(1-\theta)} + 
\frac{1}{\theta \phi} + \frac{1}{(1-\phi)}
\;,
\end{equation}
and

\begin{equation}
x^{\prime}(\alpha,\beta,\gamma,\theta,\phi)=
\frac {\gamma}{\phi(1-\theta)}(1-x(\alpha,\beta))
\;.
\end{equation}

Integrating over the parameters and expanding one gets

\begin{eqnarray}
\delta^4 N\int_p \frac {\Sigma_{9}(0)}{(p^2 +\eta^2)^2}&=&-
\delta^4 \frac {u^4}{\eta^3} \frac {N(N+2)^2}{(18)^2} 
\frac {[4.1906 \times 10^{-5}]}{(8\pi)^2} \left [ \frac {1}{\epsilon}
+ 10 \ln \left ( \frac{M}{\eta} \right ) -6.40439 \right ] 
\nonumber \\
&+&  \delta^4 \frac {u^4}{\eta^3} \frac {N(N+2)^2}{(18)^2} 
\frac {[4.1906 \times 10^{-5}]}{(8\pi)^2} 
\left [ \frac {1}{\epsilon}
+ 10 \ln \left ( \frac{M}{\eta} \right ) -6.17383 \right ]
\;,
\end{eqnarray}
which gives the finite, scale independent result

\begin{equation}
\delta^4 N\int_p \frac {\Sigma_{9}(0)}{(p^2 +\eta^2)^2}=
\delta^4 \frac {u^4}{\eta^3} \frac {N(N+2)^2}{(18)^2} 
\frac {[4.1906 \times 10^{-5}]}{(8\pi)^2}[0.23056]
\;.
\end{equation}

The final contribution comes from the last diagram of {}Fig. 3 and reads

\begin{equation}
-N\int_{p} \frac {[\Sigma_{11}(p) - \Sigma_{11}(0)]}{(p^2 +\eta^2)^2} =-
\delta^4 \frac {u^4}{\eta^3}\frac{N}{(4\pi)^{7}}
\frac{\left( 44 + 32 N + 5 N^2  \right) }{324}
[2.37741]  \;.
\end{equation}
The first contribution to this result follows from

\begin{eqnarray}
-\delta^4 N\int_{p} \frac {\Sigma_{11}(p) }{(p^2 +\eta^2)^2}&=&
+ \delta^4 \frac {u^4}{\eta^3}
\frac{N}{(4\pi)^{15/2}}
\frac{\left( 44 + 32 N + 5N^2  \right) }{324}
\Gamma(3/2+5\epsilon)\left (
\frac {M^2 \exp \gamma_E}{\eta^2} \right )^{5\epsilon}
\nonumber \\
&\times& \int_0^1 d\alpha d\beta d\gamma d\theta d\phi 
d\chi d\xi \frac{z(\alpha)z(\beta)
z(\theta)z(\phi)z(\chi) z(\gamma) z(\xi) z(\phi,\theta)
\Lambda^{-1-4 \epsilon}}
{[\gamma+\frac{\Xi}{\Lambda}(1-\gamma)]^{3/2+5\epsilon}}
\;,
\label{2luasp}
\end{eqnarray}
where

\begin{equation}
z(\alpha)=[\alpha(1-\alpha)]^{-1/2-\epsilon}  \;,
\end{equation}

\begin{equation}
z(\beta)=[\beta(1-\beta)]^{-1/2-\epsilon}  \;,
\end{equation}

\begin{equation}
z(\phi)= \phi^{-1-2 \epsilon} (1-\phi)^{-1/2+\epsilon} \;,
\end{equation}

\begin{equation}
z(\theta)=(1-\theta)^{-1/2-\epsilon}   \;,
\end{equation}

\begin{equation}
z(\chi)= (1-\chi)^{1/2+3\epsilon }  \;,
\end{equation}

\begin{equation}
z(\gamma)= \gamma (1-\gamma)^{4\epsilon }  \;,
\end{equation}

\begin{equation}
z(\xi)= \xi^{2 \epsilon} (1-\xi)^{-1/2+\epsilon }  \;,
\end{equation}

\begin{equation}
z(\phi,\theta) =\left[1-\phi (1-\theta)\right]^{-1-2\epsilon} \;,
\end{equation}

\begin{equation}
\Lambda= \frac{\theta^2 \xi (1-\chi)}{[1-\phi (1-\theta)]^2}+
\chi - \left[ \chi + \frac{\theta \xi (1-\chi)}{1-\phi (1-\theta)}\right]^2
+ \Theta \xi(1-\chi)\;,
\end{equation}

\begin{equation}
\Xi = \frac{1-\chi-\xi(1-\chi)}{\beta(1-\beta)} + \chi +
\Phi \xi (1-\chi)\;,
\end{equation}
with

\begin{equation}
\Theta = \frac{\theta(1-\theta)}{\phi (1-\theta)[1-\phi(1-\theta)]}
+1 + \frac{2 \theta}{1-\phi(1-\theta)} -
\frac{\left[1+\theta-\phi(1-\theta)\right]^2}{
\left[1-\phi (1-\theta)\right]^2}\;,
\end{equation}
and

\begin{equation}
\Phi = \frac{1-\theta-\phi(1-\theta)}{\alpha(1-\alpha) \phi(1-\theta)
[1-\phi(1-\theta)]} + \frac{\theta+ \phi(1-\theta)}{
\phi(1-\theta)[1-\phi(1-\theta)]}\;.
\end{equation}

Integrating, one obtains

\begin{eqnarray}
-\delta^4 N\int_{p} \frac {\Sigma_{11}(p) }{(p^2 +\eta^2)^2}&=&
\delta^4 \frac {u^4}{\eta^3}
\frac{N}{(4\pi)^{7}}
\frac{\left( 44 + 32 N + 5N^2  \right) }{324} [6.12476]  \;\;\;.
\end{eqnarray}

The $p=0$ case is given by
\begin{eqnarray}
\delta^4 N\int_{p} \frac {\Sigma_{11}(0) }{(p^2 +\eta^2)^2}&=&
- \delta^4 \frac {u^4}{\eta^3}\frac{N}
{(4\pi)^{15/2}}
\frac{\left( 44 + 32 N + 5N^2  \right) }{324}
\Gamma(1+4\epsilon)  \Gamma(1/2+\epsilon) \left (
\frac {M^2 \exp \gamma_E}{\eta^2} \right )^{4\epsilon}
\nonumber \\
&\times& \int_0^1 d\alpha d\beta d\xi d\theta d\phi d\chi 
\frac{z(\alpha)z(\beta)
z(\xi)z(\theta)z(\phi)z(\chi) z(\phi,\theta)}{\Xi^{1+4\epsilon}} 
\;,
\label{2luas0}
\end{eqnarray}
with the same notation as used in Eq. (\ref{2luasp}).
Integrating Eq. (\ref{2luas0}) one obtains

\begin{eqnarray}
\delta^4 N\int_{p} \frac {\Sigma_{11}(0) }{(p^2 +\eta^2)^2}&=&
-\delta^4 \frac {u^4}{\eta^3}
\frac{N}{(4\pi)^{7}}
\frac{\left( 44 + 32 N + 5N^2  \right) }{324} [8.50217]
\;.
\end{eqnarray}

%\end{references}

\newpage

\begin{table}
\begin{center}
\caption{Comparision of the results for $c_1$ as obtained from   
different methods (see text) and at different orders of approximation.}
\begin{tabular}{ccccccc}
\hline\hline
${\rm MCLS}\;\;\;\;$ & $1/N\; ({\rm LO})\;\;\;\;$  &  $1/N \;{\rm NLO}\;\;\;\;$ & 
${\rm SCR}\;\;\;\;$ &
${\cal O}(\delta^2)\;\;\;\;$ & ${\cal O}(\delta^3)\;\;\;\;$ & 
$ {\cal O}(\delta^4)$ \\
\hline
$\sim 1.30$   &  2.33 & 1.71   & 2.90  & 3.06  & 2.45 & 1.48 \\
\hline\hline
\end{tabular}
\end{center}
\end{table}

\newpage

\begin{figure}[c] 
\epsfysize=18cm  
{\centerline{\epsfbox{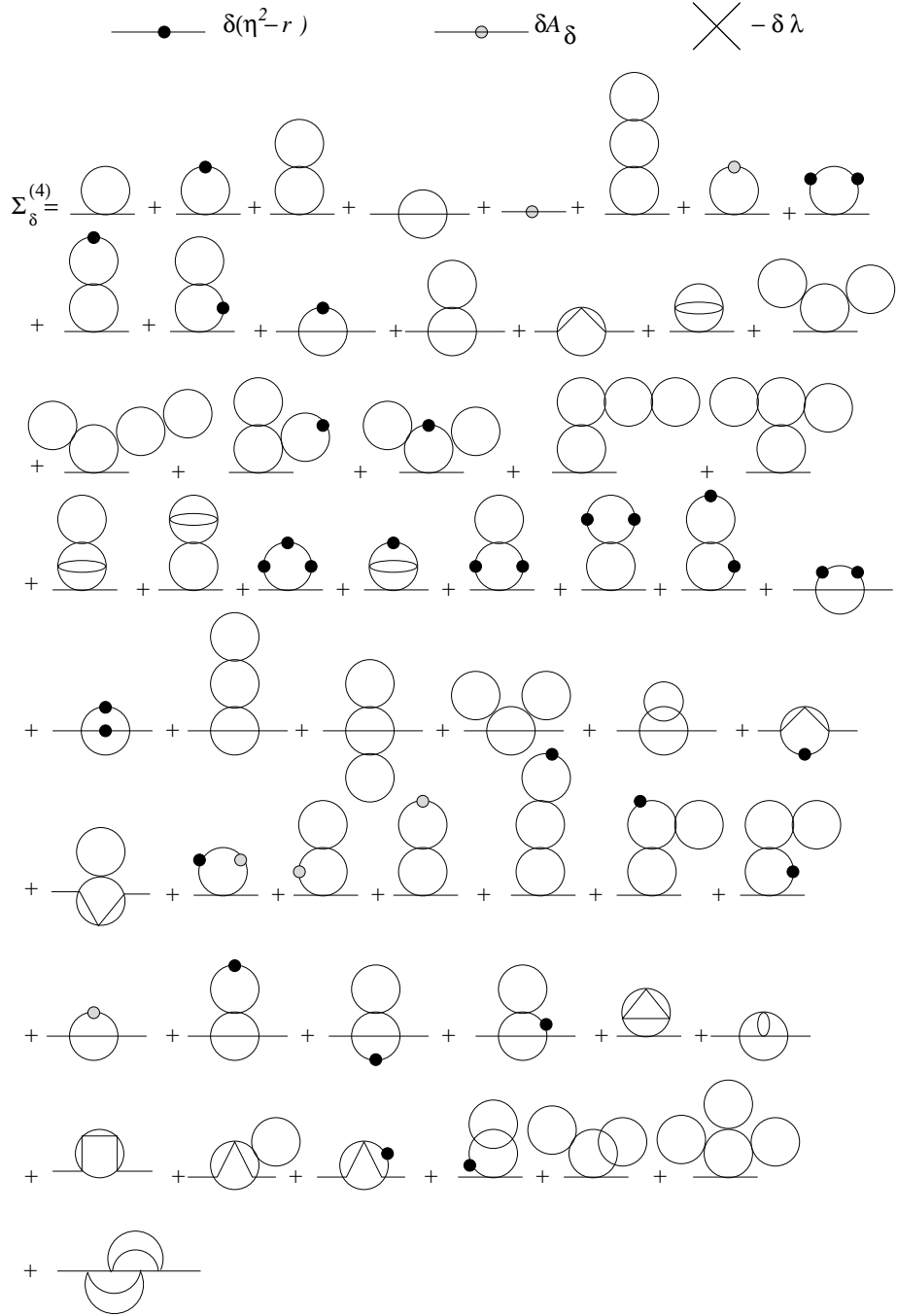}}}
\caption{Vertices (top) and diagrams contributing to the 
self-energy $\Sigma_\delta$ up to order $\delta^4$.}
%\vspace{1cm} 
 
\end{figure}

\begin{figure}[c] 
\epsfysize=7cm  
{\centerline{\epsfbox{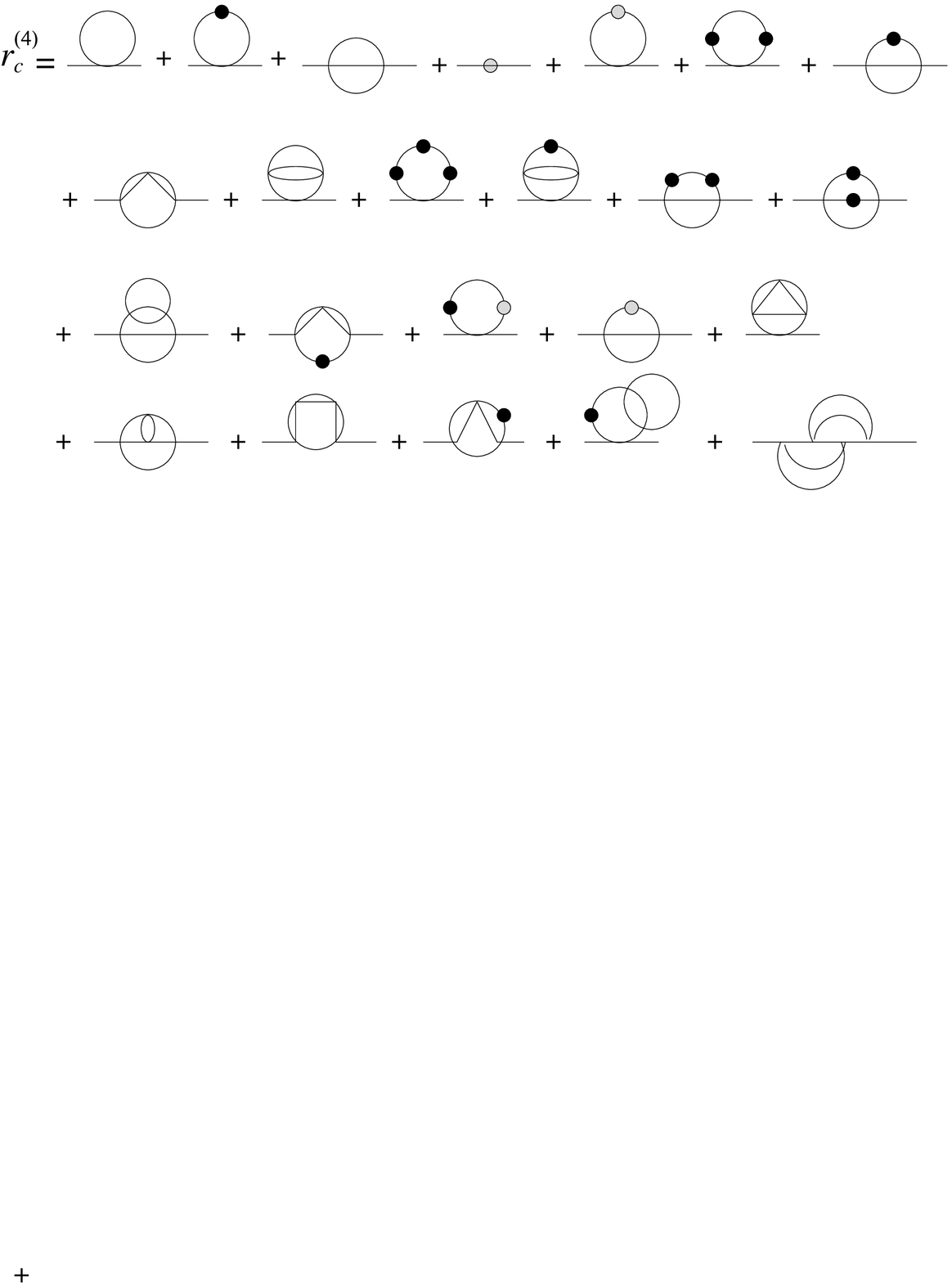}}}
\caption{The diagrams effectively contributing to $r_c$ up
to order $\delta^4$.
The black dot now represents only $\delta \eta^2$ insertions.}
%\vspace{1cm} 
 
\end{figure}

\begin{figure}[c] 
\epsfysize=6cm  
{\centerline{\epsfbox{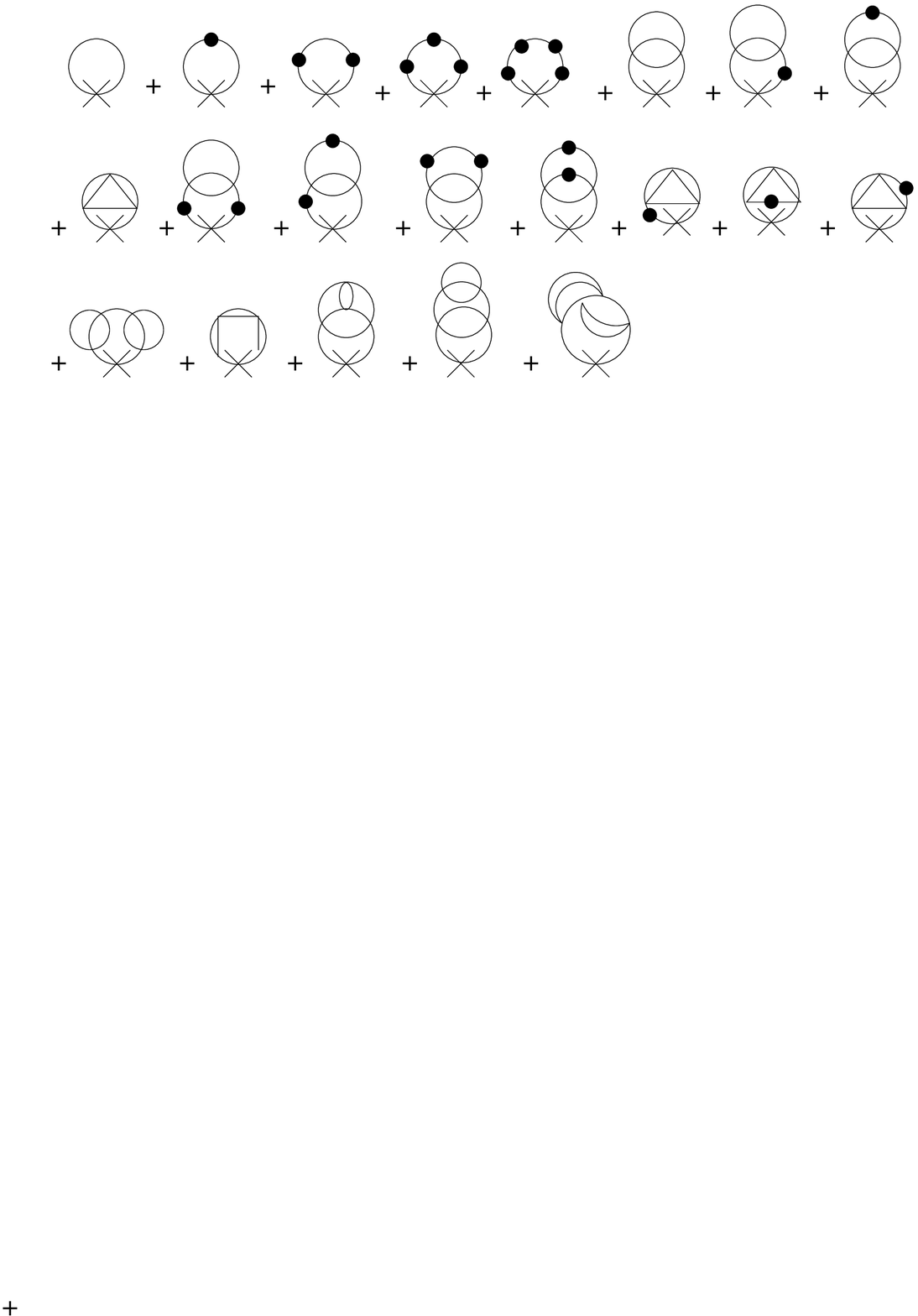}}}
\caption{All diagrams contributing to the two-point
function $\langle \phi^2 \rangle_\delta$, up to order $\delta^4$,
at the critical point. Again, the black dot represents here only the
$\delta \eta^2$ insertions.}
%\vspace{1cm} 
 
\end{figure}

\end{document}